\documentclass[superscriptaddress,twocolumn,showpacs,a4paper,longbibliography,amssymb,amsmath,nobibnotes,aps,prd,showkeys,nofootinbib,notitlepage,floatfix]{revtex4-2}
\pdfoutput=1
\usepackage{graphicx,subfigure,bm,color,psfrag,hyperref}
\usepackage{amsfonts}
\usepackage{lipsum}
\usepackage{mathtools}
\usepackage{verbatim}
\usepackage[normalem]{ulem}
\usepackage[dvipsnames]{xcolor}
\hypersetup{colorlinks,linkcolor={blue},citecolor={red},urlcolor={cyan}}

\begin{document}

\title{Beyond dynamical dark energy: the role of dark sector interactions after 
DESI DR2}

\author{Weiqiang Yang}
\email{d11102004@163.com}
\affiliation{Department of Physics, Liaoning Normal University, Dalian, 116029, 
People's Republic of China}

\author{Sibo Zhang}
\email{sbzhang02@163.com}
\affiliation{Department of Physics, Liaoning Normal University, Dalian, 116029, 
People's Republic of China}

\author{Supriya Pan}
\email{supriya.maths@presiuniv.ac.in}
\affiliation{Department of Mathematics, Presidency University, 86/1 College 
Street,  Kolkata 700073,  India}
\affiliation{Institute of Systems Science, Durban University of Technology, 
Durban 4000, Republic of South Africa}

\author{Andronikos Paliathanasis}
\email{anpaliat@phys.uoa.gr}
\affiliation{Institute of Systems Science, Durban University of Technology, 
Durban 4000,
South Africa}
\affiliation{Centre for Space Research, North-West University, Potchefstroom 
2520, South Africa. }
\affiliation{Centro de Investigaci\'on, Innovaci\'on y Creaci\'on (CIIC), 
Universidad Cat\'olica de Temuco, Temuco, Chile}
\affiliation{Departamento de Ciencias Matem\'{a}ticas y F\'{\i}sicas,  Facultad 
de Ingeniería, Universidad Cat\'olica de Temuco, Temuco, Chile}
\affiliation{National Institute for Theoretical and Computational Sciences 
(NITheCS), South Africa }

\author{Emmanuel N. Saridakis} 
\email{msaridak@noa.gr}
\affiliation{National Observatory of Athens, Lofos Nymfon 11852, Greece}
\affiliation{CAS Key Laboratory for Research in Galaxies and Cosmology, School 
of Astronomy and Space Science, \\
University of Science and Technology of China, Hefei 230026, China}
\affiliation{Departamento de Matem\'{a}ticas, Universidad Cat\'{o}lica del 
Norte, Avda. Angamos 0610, Casilla 1280, Antofagasta, Chile}




\begin{abstract}

Recent DESI DR2 observations have renewed interest in extensions of the
$\Lambda$CDM cosmological model, particularly through indications of a
time-varying dark energy equation of state. In this work, we investigate
whether such deviations may also involve interactions within the dark sector.
We consider an interacting dark energy scenario in which the dark matter density
evolves as $\rho_{\rm dm}\propto a^{-3+\delta}$, with the constant $\delta$
quantifying the interaction strength, and allow the dark energy equation of
state to be either constant but different from $-1$, or dynamically evolving
through the CPL parametrization. The models are constrained using Planck CMB
data, DESI DR2 BAO measurements, and three Type Ia supernova compilations:
PantheonPlus, Union3, and DES-Dovekie. 
For the constant equation-of-state case, the inclusion of DESI and supernova
data leads to a preference for a small negative interaction parameter, with a
significance above $2\sigma$. When dynamical dark energy is allowed, the
evidence for interaction becomes weak, while the data favor a
quintessence-like evolving dark energy component. In both scenarios, Bayesian
model comparison still favors $\Lambda$CDM. Our results show that the inferred
role of dark-sector interactions depends strongly on the nature of dark energy,
highlighting the importance of jointly testing dark energy dynamics and
interactions in the DESI era.

\end{abstract}

\maketitle
\section{Introduction}

The standard $\Lambda$-Cold Dark Matter ($\Lambda$CDM) cosmological model has
provided a remarkably successful description of a wide range of cosmological
observations. Nevertheless, the physical nature of its two dominant dark
components, dark matter (DM) and dark energy (DE), remains one of the most
fundamental open questions in modern cosmology. Hence, there is no
fundamental reason requiring the dark components to be completely decoupled.
This motivates interacting dark energy (IDE) scenarios, in which DM and DE are
allowed to exchange energy and/or momentum through a non-gravitational
interaction. Such an interaction can modify both the background expansion
history and the evolution of cosmological perturbations, leaving potentially
observable signatures in the cosmic microwave background (CMB), large-scale
structure, and distance measurements.

The possibility of an interaction in the dark sector has been extensively
investigated using both parametric approaches
~\cite{Amendola:1999er,Amendola:1999qq,Cai:2004dk,Barrow:2006hia,Guo:2004xx,
Wang:2005ph,Chimento:2003iea, 
Amendola:2006dg,  Majerotto:2009np,
Jamil:2009eb,Koshelev:2010umw,Chen:2011cy, 
Yang:2014gza, 
Faraoni:2014vra, Wang:2014xca,vandeBruck:2015ida,Pan:2012ki, 
vandeBruck:2016jgg,Nunes:2016dlj,Wang:2016lxa,
Yang:2016evp,vandeBruck:2016hpz,Yang:2017zjs,Pan:2017ent,Mifsud:2017fsy,
Xu:2017rfo,Yang:2017ccc,VanDeBruck:2017mua,Yang:2017zjs,Yang:2018euj,
Yang:2018xlt, Pan:2019jqh, 
Li:2019loh, 
Pan:2019gop,Yang:2019bpr,Yang:2019vni,Khyllep:2021wjd,
 Barrow:2019jlm,vonMarttens:2019ixw,Yang:2019uog,
DiValentino:2019ffd,DiValentino:2019jae,Gomez-Valent:2020mqn,
Gomez-Valent:2022bku,Pan:2022qrr, Nunes:2022bhn, 
Zhai:2023yny,Hoerning:2023hks,Forconi:2023hsj,Teixeira:2023zjt,Li:2023gtu, 
Halder:2024aan,Wang:2024vmw, 
Giare:2024ytc,Sabogal:2024yha,Li:2024qso, 
Ghedini:2024mdu,Li:2025owk,
Wang:2025znm, Yang:2025boq,
Cruickshank:2025iig,Feng:2025mlo, 
Goh:2024exx,Zhai:2025hfi,Nari:2025thx,
You:2025uon,vanderWesthuizen:2025iam,Postolak:2025qmv,Zhang:2025dwu,
Silva:2025bnn,Wu:2025vrl,Petri:2025swg,Li:2025muv,Paliathanasis:2026ymi,
Dai:2026pvx, 
Meng:2026lzz,Li:2026xaz,Figueruelo:2026eis,Sahlu:2026bsa, Gomez-Valent:2026ept,
Paliathanasis:2026acn,Zhai:2026uwr,Wang:2026vqw,Villalobos:2026zhz,
Antusch:2026ldp,Maier:2026cdz,Ambelu:2026mkd,Wang:2026wrk,Li:2026ldf,
BarrosoVarela:2026zij}
and non-parametric approaches
~\cite{Yang:2015tzc,Cai:2017yww,vonMarttens:2020apn,Bonilla:2021dql,Mukherjee:2021ggf, 
Escamilla:2023shf,Abedin:2025yru,Li:2025ula,You:2025uon}. These studies have demonstrated that interactions in the dark sector can provide
interesting phenomenological possibilities, including potential alleviation of 
the coincidence problem \cite{Amendola:2000uh,Billyard:2000bh,
Zimdahl:2001ar,Wetterich:1994bg,Cai:2004dk, delCampo:2008sr},  and of the 
existing cosmological tensions, such as the $H_0$ one
  \cite{DiValentino:2015ola,Kumar:2017dnp,
Khosravi:2017hfi,Mortsell:2018mfj, 
Yang:2018qmz,Martinelli:2019dau,
Agrawal:2019lmo, 
Yang:2019nhz,Giare:2024smz}, and the $\sigma_8$  one
\cite{Pourtsidou:2016ico,An:2017crg,DiValentino:2018gcu,Kumar:2019wfs,
Khoury:2025txd},  as well as the presence of
accelerating scaling solution 
\cite{Amendola:1999er,Amendola:1999qq,Cai:2004dk,Barrow:2006hia,Guo:2004xx,Chimento:2003iea,Chen:2008pz,
Khyllep:2021wjd,Halder:2024aan,Halder:2025ytq,Postolak:2025qmv,Paliathanasis:2026ymi}.

Within the parametric framework of interacting cosmologies, the interaction
function is usually introduced phenomenologically and the resulting cosmological
model is confronted with observations. Although several choices of interaction
forms have been proposed, often motivated by field-theoretical considerations,
a unique fundamental action describing interacting dark energy has not yet been
established. Therefore, an alternative and particularly useful approach is to
characterize the interaction through its impact on the evolution of dark matter
itself. In the absence of interactions, pressureless dark matter follows the
standard scaling law
$
\rho_{\rm dm}\propto a^{-3}$,
where $a$ is the scale factor. However, an exchange of energy with dark energy
naturally modifies this evolution. A simple and widely studied phenomenological
parametrization is therefore given by
~\cite{Wang:2004cp,Amendola:2006dg,Costa:2009mv,Wei:2010uh,Nunes:2014qoa,
Wang:2014iua,Nunes:2016dlj,Kumar:2016zpg,Pan:2025qwy}
\begin{eqnarray}\label{scaling-DM}
\rho_{\rm dm}=\rho_{\rm dm,0}a^{-3+\delta},
\end{eqnarray}
where $\delta$ quantifies the departure from the standard evolution and may be
either constant or time dependent. Positive values of $\delta$ correspond to a
slower dilution of dark matter compared with the standard $a^{-3}$ evolution,
whereas negative values indicate a faster dilution. Consequently, deviations
from $\delta=0$ provide a direct phenomenological measure of a possible
interaction between the dark components. This approach has received significant
attention in the literature.

The recent second data release of the Dark Energy Spectroscopic Instrument
(DESI) has renewed interest in extensions of $\Lambda$CDM, particularly due to
indications for a time-varying dark energy equation of state. These results
raise an important question: if the dark energy sector is dynamical, is this
sufficient to explain the possible deviations from $\Lambda$CDM, or could
dark-sector interactions provide an additional contribution? Answering this
question requires studying the interaction framework beyond the usual
assumption of a cosmological constant.

After the DESI DR2 results, the above dark matter scaling scenario was
investigated in~\cite{Pan:2025qwy}, assuming a vacuum dark energy component with
a fixed equation of state $w=-1$. In the present work, we extend this analysis 
by
relaxing this assumption and allowing the dark energy equation of state to be
either constant but different from $-1$, or dynamically evolving according to
the CPL parametrization. This extension is essential because the inferred
strength and even the preferred direction of the dark-sector interaction can
depend on the intrinsic properties of dark energy. Therefore, our analysis
provides a direct investigation of the interplay between dynamical dark energy
and dark-sector interactions in the DESI era.

We constrain the resulting scenarios using a comprehensive combination of 
current
cosmological observations, including Planck CMB data, DESI DR2 BAO measurements,
and three independent Type Ia supernova compilations, namely PantheonPlus,
Union3, and DES-Dovekie. By combining background and perturbation analyses, we
investigate how the interaction parameter and the dark energy equation of state
jointly affect cosmological observables. Our results reveal that the role of
dark-sector interactions is strongly connected with the assumed nature of dark
energy, providing new insight into whether the DESI-era deviations from
$\Lambda$CDM are driven by dynamical dark energy, dark-sector interactions, or a
combination of both.

The paper is organized as follows. In Sec.~\ref{sec-2}, we describe the
theoretical framework of the interacting scenario. Sec.~\ref{se-data} presents
the observational datasets, numerical implementation, and statistical
methodology. In Sec.~\ref{sec-results}, we discuss the observational constraints
and the implications of the different interacting scenarios. Finally,
Sec.~\ref{sec-summary} summarizes our conclusions.

\section{Dark sector interaction framework}
\label{sec-2}

In this section, we present the theoretical framework of the interacting dark
energy scenarios considered in this work. We focus on a phenomenological
description in which the interaction between dark matter and dark energy is
encoded through a modification of the standard dark matter dilution law. Such
an interaction affects both the background expansion history and the evolution
of cosmological perturbations, and therefore both aspects must be consistently
included when confronting the model with observations. We first discuss the
background dynamics of the interacting scenarios, including the different
choices adopted for the dark energy equation of state. We then describe the
evolution of linear perturbations, which determines the impact of the
interaction on cosmic microwave background anisotropies and large-scale
structure observables.

\subsection{Background evolution}

We consider a spatially flat Friedmann-Lema\^{i}tre-Robertson-Walker (FLRW) 
universe
and assume that the gravitational sector is described by general relativity. 
The cosmic inventory consists of baryons, radiation, neutrinos, pressureless 
dark matter (DM),
and dark energy (DE). While the total energy density is conserved, we allow for 
a
non-gravitational energy exchange between the dark components, so that DM and 
DE 
do not
evolve independently.

In the presence of such an interaction, the DM density no longer necessarily 
follows
the standard scaling law $ \rho_{\rm dm}\propto a^{-3}$. Instead, we assume the
phenomenological form given by eqn.~(\ref{scaling-DM}), which leads to the 
coupled
continuity equations
\begin{align}\label{coupled-eqns}
\dot{\rho}_{\rm dm} + 3 H \rho_{\rm dm} =  - 
\dot{\rho}_{\rm de} - 3 H (1+w (a)) \rho_{\rm de} = - Q,
\end{align}
where $\rho_{\rm dm}$ and $\rho_{\rm de}$ denote the energy densities of dark 
matter
and dark energy, respectively, $w(a)$ is the dark energy equation-of-state 
parameter,
and $Q$ describes the energy transfer between the two dark fluids.

For a constant interaction parameter,
$\delta=\delta_0$,\footnote{The case of a dynamical $\delta$ will be explored 
in 
a
companion work including a detailed observational analysis.}
the interaction function becomes
\begin{eqnarray}\label{model-Q}
Q = -\delta_0 H \rho_{\rm dm},
\end{eqnarray}
which corresponds to one of the most widely studied phenomenological interacting
dark energy scenarios~\cite{Wang:2004cp,Amendola:2006dg,He:2008tn,Costa:2009mv,
Wei:2010uh,Nunes:2014qoa,Wang:2014iua,Nunes:2016dlj,Kumar:2016zpg,Pan:2025qwy,
Holanda:2019sod,Bora:2021uxq,Bora:2021iww}.

The background expansion is then governed by the Friedmann equation
\begin{align}\label{Hubble-expansion}
\left(\frac{H}{H_0} \right)^2 =
\Omega_{b}a^{-3}+ \Omega_{r}a^{-4}
+\frac{\rho_{\nu}}{\rho_{\rm ct,0}}
+\Omega_{\rm dm}a^{-3+\delta_0}
+\frac{\rho_{\rm de}}{\rho_{\rm ct,0}},
\end{align}
where $\Omega_b$, $\Omega_r$, and $\Omega_{\rm dm}$ are the present-day density
parameters of baryons, radiation, and dark matter, respectively, while
$\rho_{\rm ct,0}$ denotes the present critical density and $\rho_\nu$ the 
neutrino
energy density.

The evolution of the dark energy density is obtained by solving the DE 
conservation
equation in (\ref{coupled-eqns}) after substituting the interaction term in
(\ref{model-Q}). Its behavior depends on the adopted form of the DE equation of 
state.
Equivalently, the interacting system can be described through effective 
equations of
state for two separately conserved fluids,
\begin{eqnarray}
&&\dot{\rho}_{\rm dm} + 3H(1+w_{\rm dm}^{\rm eff})\rho_{\rm dm}=0,\\
&&\dot{\rho}_{\rm de} + 3H(1+w_{\rm de}^{\rm eff})\rho_{\rm de}=0,
\end{eqnarray}
where
\begin{eqnarray}
&&w_{\rm dm}^{\rm eff}\equiv \frac{Q}{3H\rho_{\rm dm}}
=-\frac{\delta_0}{3},\\
&&w_{\rm de}^{\rm eff}\equiv w(a)-\frac{Q}{3H\rho_{\rm de}}
=w(a)+\frac{\delta_0}{3}\frac{\rho_{\rm dm}}{\rho_{\rm de}}.
\end{eqnarray}
These relations show that the sign of $\delta_0$ determines the direction of the
energy transfer and modifies the effective behavior of both dark components.
Furthermore, the effective dark energy equation of state depends not only on the
interaction strength but also on the intrinsic DE equation of state $w(a)$.

In this work we consider two choices for the DE equation of state:

\begin{enumerate}

\item We first consider a constant equation of state,
\begin{eqnarray}
w(a)=w_0,
\end{eqnarray}
where $w_0$ denotes its present-day value. The corresponding interacting model 
is
denoted as {\bf ``IDE$+w$''}.

\item We also consider a dynamical dark energy scenario described by the
Chevallier-Polarski-Linder (CPL) parametrization
\cite{Chevallier:2000qy,Linder:2002et},
\begin{eqnarray}
w(a)=w_0+w_a(1-a),
\end{eqnarray}
where $w_a$ quantifies the time evolution of the equation of state. This 
scenario is
denoted as {\bf ``IDE$+w_0w_a$''}.

\end{enumerate}

These two choices allow us to investigate whether dark sector interactions 
provide
additional degrees of freedom beyond constant and dynamical dark energy 
descriptions.

\subsection{Linear perturbation evolution}

To consistently confront the interacting scenarios with cosmological 
observations, it is
necessary to account for their impact at the level of linear perturbations. The 
dark
sector interaction modifies not only the background evolution but also the 
growth of
density fluctuations and the corresponding observational signatures. We 
therefore
consider the perturbed FLRW metric in the synchronous gauge, which takes the 
form
\cite{Ma:1995ey}
\begin{eqnarray}
ds^2 &=& a^2(\tau) \Big[ 
- d\tau^2 + (\delta_{ij} + h_{ij})dx^{i} dx^{j} \, \Big],
\end{eqnarray}
where $\tau$ denotes the conformal time, $a(\tau)$ is the scale factor, while
$\delta_{ij}$ and $h_{ij}$ represent the background and perturbed spatial metric
components, respectively.

The evolution of the density contrasts and velocity divergences of the 
interacting
dark matter and dark energy fluids is governed by the coupled perturbation 
equations
\cite{Ma:1995ey}
	\begin{eqnarray}
		&&\delta _{\rm de}^{\prime } =-[1\!+\!w (a)]\left(\! \theta _{\rm 
de}+\frac{h^{\prime }}{2}%
		\right)-\delta_0\left(\!\theta+\frac{h^\prime}{2}\right) 
		-3\mathcal{H}w (a)^{\prime}\frac{\theta_x}{k^{2}}
		\nonumber\\
&& \ \ \ \ \ \ -3\mathcal{H}[c_{s, {\rm de}}^{2}-w (a)]\left[ \delta _{\rm 
de}+3\mathcal{H}%
		[1\!+\!w (a)]\frac{\theta _x}{k^{2}}\right] \nonumber \\
	 &&
	\ \ \ \ \ \ 	-3\mathcal{H}\delta_0\frac{\rho_{\rm dm}}{\rho_{\rm 
de}} 
\left\{ \delta_{\rm dm}-\delta_{\rm de}+3\mathcal{H} \left[c_{s,{\rm 
de}}^2\!-\!w (a) 
\right]\frac{\theta_{\rm de}}{k^2} 
\right\}, \\
		&&\theta _{\rm de}^{\prime } =-\mathcal{H}(1-3c_{s,{\rm de}}^{2})\theta 
_{\rm de}+\frac{%
			c_{s,{\rm de}}^{2}}{[1\!+\!w (a)]}k^{2}\delta _{\rm 
de}\nonumber\\
&&
	\ \ \ \ \ \ \ \ 
+3\mathcal{H}\delta_0\frac{\rho_c}{\rho_{\rm de}}\left[ \frac{%
			(1\!+\!c_{s, {\rm de}}^{2})\theta _{\rm de}}{1\!+\!w (a)}\right], \\
		&&\delta _{\rm dm}^{\prime } =-\frac{h^{\prime 
}}{2}+\delta_0\left(\theta_{\rm dm}+\frac{h^\prime}{2}\right),\\
        &&  \theta_{\rm dm}^\prime = - \mathcal{H} \theta_{\rm 
dm}, 
	\end{eqnarray}
where a prime denotes differentiation with respect to the conformal time $\tau$.
The quantities $\delta_{\rm de}$ ($\delta_{\rm dm}$) and
$\theta_{\rm de}$ ($\theta_{\rm dm}$) represent, respectively, the density 
contrasts
and velocity divergences (in Fourier space) of the dark energy (dark matter) 
fluids.
Furthermore, $h$ denotes the trace of the metric perturbation $h_{ij}$,
$\mathcal{H}=a'(\tau)/a(\tau)$ is the conformal Hubble parameter, and
$c_{s,\rm de}^{2}$ is the physical (rest-frame) sound speed squared of the dark
energy fluid, which is fixed to unity throughout this work.

The above equations determine the evolution of cosmological perturbations in the
interacting scenarios and allow us to consistently evaluate their impact on
large-scale structure and cosmic microwave background observables.

 \section{Observational datasets and statistical analysis}
\label{se-data}

In this section we describe the observational datasets, numerical 
implementation, and
statistical methodology adopted to constrain the interacting dark energy 
scenarios.
As discussed above, we consider two extensions of $\Lambda$CDM: one in which 
the 
dark
energy equation of state is constant but not necessarily equal to $-1$, and one 
in
which it follows the dynamical CPL parametrization.

For the constant equation-of-state case, the six standard $\Lambda$CDM 
parameters are
supplemented by the interaction parameter $\delta_0$ and the DE 
equation-of-state
parameter $w_0$, leading to an eight-dimensional parameter space,
\begin{align}
\Bigl\{\Omega_{\rm b}h^2, \Omega_{\rm c}h^2, 100\theta_{\rm MC},
\tau, n_s, \ln(10^{10}A_s), \delta_0, w_0 \Bigr\}.
\end{align}
For the dynamical dark energy case described by the CPL parametrization, an 
additional
parameter $w_a$ is introduced, resulting in a nine-dimensional parameter space,
\begin{align}
\Bigl\{\Omega_{\rm b}h^2, \Omega_{\rm c}h^2, 100\theta_{\rm MC},
\tau, n_s, \ln(10^{10}A_s), \delta_0, w_0, w_a \Bigr\}.
\end{align}

The common parameters have their standard meanings:
$\Omega_b h^2$ and $\Omega_c h^2$ denote the physical baryon and cold dark
matter densities, respectively; $\theta_{\rm MC}$ is the angular size of the
sound horizon at recombination; $\tau$ is the optical depth to reionization;
$n_s$ is the scalar spectral index; and $A_s$ represents the amplitude of the
primordial scalar perturbation spectrum. The prior ranges adopted for all free
parameters in the statistical analysis are summarized in 
Table~\ref{tab:priors}.

\begin{table}[!ht]
\caption{Prior ranges of the free cosmological parameters considered in the
statistical analysis of the interacting dark energy scenarios.}
	\begin{center}
		\renewcommand{\arraystretch}{1.4}
		\begin{tabular}{|c@{\hspace{1 cm}}|@{\hspace{1 cm}} c|}
			\hline
			\textbf{Parameter}           & \textbf{Prior}\\
			\hline\hline
			$\Omega_{b} h^2$             & $[0.005,0.1]$ \\
			$\Omega_{\rm dm} h^2$         & $[0.001,0.99]$\\
			$\tau$                       & $[0.01,0.8]$\\
			$n_s$                        & $[0.8, 1.2]$\\
                $100\theta_{MC}$             & $[0.5,10]$\\ 
			$\log[10^{10}A_{s}]$         & $[1.61,3.91]$ \\
			
			$\delta_0$                   & $[-1, 1]$ \\	
			$w_0$						 & $[-3, 3]$ \\
            $w_a$                         & $[-3, 3]$\\
			\hline
		\end{tabular}
	\end{center}
	\label{tab:priors}
\end{table}

\begingroup
\squeezetable

\begin{center}  
	\begin{table*}
	\caption{68\% and 95\% confidence level constraints on the free and derived
cosmological parameters of the {\bf IDE$+w$} scenario for the CMB, CMB+DESI,
CMB+DESI+PantheonPlus, CMB+DESI+Union3, and CMB+DESI+DES-Dovekie dataset
combinations.}
		\begin{tabular}{cccccc}
			\hline\hline
			Parameters & CMB & CMB+DESI & CMB+DESI+PantheonPlus & 
CMB+DESI+Union3 & CMB+DESI+DES-Dovekie  \\ \hline
			
			$\Omega_\mathrm{b} h^2$ & 
$0.02241_{-0.00016-0.00032}^{+0.00017+0.00031}$ & 
$0.02234_{-0.00017-0.00033}^{+0.00017+0.00032}$ & 
$0.02233_{-0.00017-0.00032}^{+0.00016+0.00031}$ & 
$0.02232_{-0.00016-0.00032}^{+0.00016+0.00032}$ & 
$0.02233_{-0.00016-0.00030}^{+0.00016+0.00031}$ \\
			
			$\Omega_\mathrm{dm} h^2$ & 
$0.12546_{-0.00369-0.00686}^{+0.00306+0.00704}$ & 
$0.11714_{-0.00105-0.00224}^{+0.00115+0.00210}$ & 
$0.11636_{-0.00095-0.00187}^{+0.00093+0.00185}$ & 
$0.11611_{-0.00102-0.00204}^{+0.00103+0.00197}$ & 
$0.11629_{-0.00096-0.00189}^{+0.00095+0.00183}$ \\
			
			$100\theta_\mathrm{MC}$ & 
$1.04018_{-0.00038-0.00191}^{+0.00052+0.00098}$ & 
$1.04093_{-0.00028-0.00056}^{+0.00028+0.00058}$ & 
$1.04097_{-0.00028-0.00057}^{+0.00029+0.00057}$ & 
$1.04101_{-0.00028-0.00053}^{+0.00028+0.00053}$ & 
$1.04099_{-0.00028-0.00055}^{+0.00028+0.00057}$ \\
			
			$\tau$ & $0.0530_{-0.0080-0.0152}^{+0.0076+0.0160}$ & 
$0.0576_{-0.0083-0.0152}^{+0.0074+0.0171}$ & 
$0.0581_{-0.0085-0.0152}^{+0.0072+0.0175}$ & 
$0.0581_{-0.0083-0.0153}^{+0.0074+0.0163}$ & 
$0.0579_{-0.0085-0.0162}^{+0.0075+0.0167}$ \\
			
			$n_\mathrm{s}$ & $0.9671_{-0.0054-0.0108}^{+0.0054+0.0103}$ & 
$0.9777_{-0.0037-0.0070}^{+0.0037+0.0071}$ & 
$0.9784_{-0.0035-0.0069}^{+0.0035+0.0067}$ & 
$0.9785_{-0.0036-0.0069}^{+0.0035+0.0072}$ & 
$0.9782_{-0.0037-0.0073}^{+0.0037+0.0071}$ \\
			
			$\ln(10^{10} A_\mathrm{s})$ & $3.053_{-0.016-0.031}^{+0.016+0.032}$ 
& $3.058_{-0.016-0.032}^{+0.016+0.034}$ & $3.059_{-0.017-0.031}^{+0.015+0.035}$ 
& $3.059_{-0.016-0.032}^{+0.016+0.034}$ & $3.059_{-0.018-0.031}^{+0.015+0.035}$ 
\\
			
			$\delta_0$ & $0.0043_{-0.0028-0.0056}^{+0.0024+0.0055}$ & 
$-0.0018_{-0.0013-0.0026}^{+0.0013+0.0025}$ & 
$-0.0025_{-0.0012-0.0024}^{+0.0012+0.0023}$ & 
$-0.0028_{-0.0013-0.0024}^{+0.0012+0.0024}$ & 
$-0.0026_{-0.0012-0.0022}^{+0.0012+0.0023}$ \\
			
			$w_0$ & $-0.920_{-0.292-0.361}^{+0.134+0.484}$ & 
$-0.998_{-0.046-0.094}^{+0.046+0.091}$ & $-0.956_{-0.027-0.052}^{+0.027+0.052}$ 
& $-0.942_{-0.033-0.064}^{+0.032+0.064}$ & 
$-0.949_{-0.026-0.049}^{+0.025+0.049}$ \\
			\hline 
            
			$\Omega_\mathrm{m}$ & $0.3976_{-0.1046-0.1442}^{+0.0411+0.2119}$ & 
$0.2971_{-0.0075-0.0146}^{+0.0076+0.0151}$ & 
$0.3028_{-0.0051-0.0099}^{+0.0051+0.0105}$ & 
$0.3048_{-0.0057-0.0113}^{+0.0056+0.0115}$ & 
$0.3040_{-0.0048-0.0093}^{+0.0048+0.0093}$ \\
			
			$\sigma_8$ & $0.776_{-0.032-0.089}^{+0.047+0.074}$ & 
$0.817_{-0.012-0.024}^{+0.013+0.024}$ & $0.812_{-0.011-0.022}^{+0.011+0.022}$ & 
$0.810_{-0.011-0.022}^{+0.011+0.022}$ & $0.811_{-0.012-0.021}^{+0.011+0.022}$ \\
			
			$H_0$ [km/s/Mpc] & $62.18_{-4.74-13.43}^{+7.15+10.39}$ & 
$68.69_{-1.01-1.94}^{+0.97+1.98}$ & $67.84_{-0.59-1.15}^{+0.59+1.13}$ & 
$67.56_{-0.69-1.33}^{+0.71+1.38}$ & $67.70_{-0.55-1.04}^{+0.53+1.06}$ \\
			
			$S_8$ & $0.882_{-0.059-0.088}^{+0.034+0.113}$ & 
$0.813_{-0.011-0.022}^{+0.011+0.022}$ & $0.815_{-0.011-0.021}^{+0.011+0.022}$ & 
$0.817_{-0.011-0.021}^{+0.011+0.022}$ & $0.816_{-0.011-0.021}^{+0.011+0.022}$ \\
			
			$r_{\rm{drag}}$ [Mpc] & $146.71_{-0.36-0.76}^{+0.40+0.72}$ & 
$147.52_{-0.23-0.45}^{+0.24+0.47}$ & $147.59_{-0.23-0.43}^{+0.23+0.44}$ & 
$147.61_{-0.24-0.45}^{+0.23+0.46}$ & $147.58_{-0.23-0.46}^{+0.24+0.46}$ \\
			
			\hline
			
			$w_0~(\sigma)$ & $0.2002$ & $0.0684$ & $1.6447$ & $1.7665$ & 
$2.0264$\\
			
			$\delta_0~(\sigma)$ & $1.6377$ & $1.4336$ & $2.1813$ & $2.3394$ & 
$2.2484$\\
			
			
			$\rm{ln}\mathcal{B}_{ij}$ & $-7.2$ & $-8.6$ & $-10.8$ & $-10.1$ & 
$-10.1$\\
		
			\hline                                                         
		\end{tabular}                                                       
			\label{table-1}                   
	\end{table*}                                     
\end{center}
\endgroup

To constrain these parameter spaces, we modify the publicly available Boltzmann
code CAMB~\cite{Lewis:1999bs,Lewis:2002ah} to incorporate the interacting dark
sector dynamics.  The parameter inference is performed using the publicly 
available
Monte Carlo sampling framework \texttt{Cobaya}~\cite{Torrado:2020dgo}, which 
provides
access to the latest cosmological likelihoods. The convergence of the Markov 
chains is
assessed using the Gelman--Rubin statistic~\cite{Gelman:1992zz}, while the
\texttt{GetDist} package~\cite{Lewis:2019xzd} is used for the analysis and
visualization of the posterior distributions.

The observational datasets employed in our analysis are summarized below.

\begin{itemize}

\item {\bf Cosmic microwave background (CMB):}
We use the Planck 2018 temperature and polarization angular power spectra
\cite{Planck:2018vyg,Planck:2019nip}, specifically the likelihood combination
\textit{plikTTTEEE+lowl+lowE}.

\item {\bf Baryon acoustic oscillations (BAO):}
We include the BAO measurements from the second data release (DR2) of the DESI 
survey
\cite{DESI:2025zgx}, which provide constraints on the expansion history and
cosmological distances. We refer to this dataset as DESI throughout the paper.

\item {\bf Type Ia supernovae (SN Ia):}
We consider three independent supernova compilations:
(i) the PantheonPlus sample~\cite{Scolnic:2021amr};
(ii) the expanded Union3 compilation containing 2087 supernovae
\cite{Rubin:2023jdq};
and (iii) the five-year Dark Energy Survey supernova analysis
(labeled DES-Dovekie)~\cite{DES:2025sig}.

\end{itemize}

We quantify the performance of the interacting scenarios using  
the Bayesian evidence relative to $\Lambda$CDM. For a 
model
$M_i$ and dataset $\mathcal{D}$, the Bayesian evidence is defined as
\begin{eqnarray}
B_i=\int \mathcal{L}(\mathcal{D}|\Theta,M_i)
\pi(\Theta|M_i)d\Theta ,
\end{eqnarray}
where $\pi(\Theta|M_i)$ denotes the prior distribution, $\Theta$ the parameter
vector, and $\mathcal{L}$ the likelihood function. Comparing with the reference 
model
$M_j=\Lambda$CDM gives the Bayes factor
\begin{eqnarray}
\mathcal{B}_{ij}=\frac{B_i}{B_j},
\end{eqnarray}
or equivalently,
\begin{eqnarray}
\ln\mathcal{B}_{ij}=\ln B_i-\ln B_j .
\end{eqnarray}
Positive values of $\ln\mathcal{B}_{ij}$ indicate a preference for the extended 
model
over $\Lambda$CDM. Following the revised Jeffreys scale
\cite{Kass:1995loi}, we classify the evidence as inconclusive for
$0\leq\ln\mathcal{B}_{ij}<1$, weak for $1\leq\ln\mathcal{B}_{ij}<2.5$,
moderate for $2.5\leq\ln\mathcal{B}_{ij}<5$, strong for
$5\leq\ln\mathcal{B}_{ij}<10$, and very strong for
$\ln\mathcal{B}_{ij}\geq10$.

The Bayesian evidence calculations are performed using
\texttt{MCEvidence}~\cite{Heavens:2017afc} together with the
\texttt{Cobaya} wrapper implemented in the \texttt{wgcosmo} repository
\cite{Giare:wgcosmo, Yang:2026kxp, Yang:2025gaz}.

\begin{figure*}
	\includegraphics[width=0.85\textwidth]{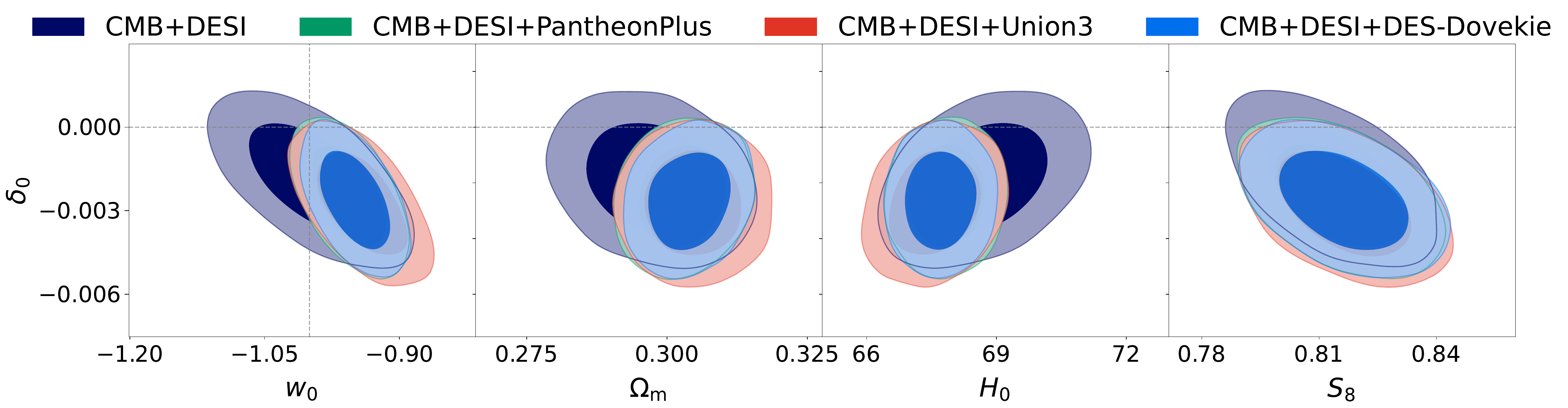}
	\includegraphics[width=0.85\textwidth]{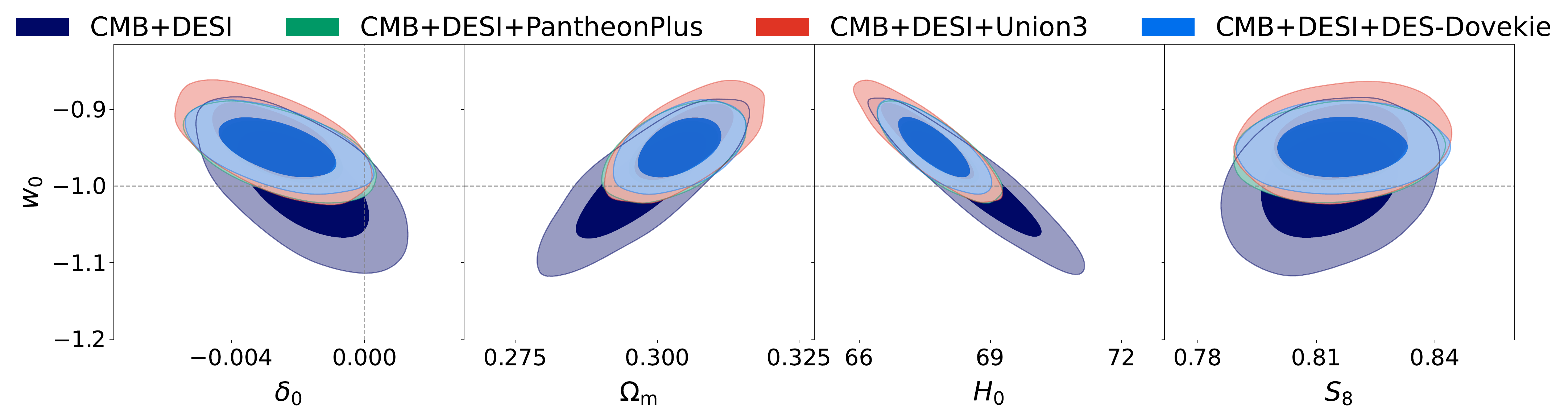}	
\caption{\textit{Marginalized one-dimensional posterior distributions and
two-dimensional confidence contours for the main cosmological parameters of
the {\bf IDE$+w$} scenario. Results are displayed for the CMB, CMB+DESI,
CMB+DESI+PantheonPlus, CMB+DESI+Union3, and CMB+DESI+DES-Dovekie dataset
combinations.}}
	\label{contour-1}
\end{figure*}
\begin{figure*}
	\centering
	\includegraphics[width=0.47\textwidth]{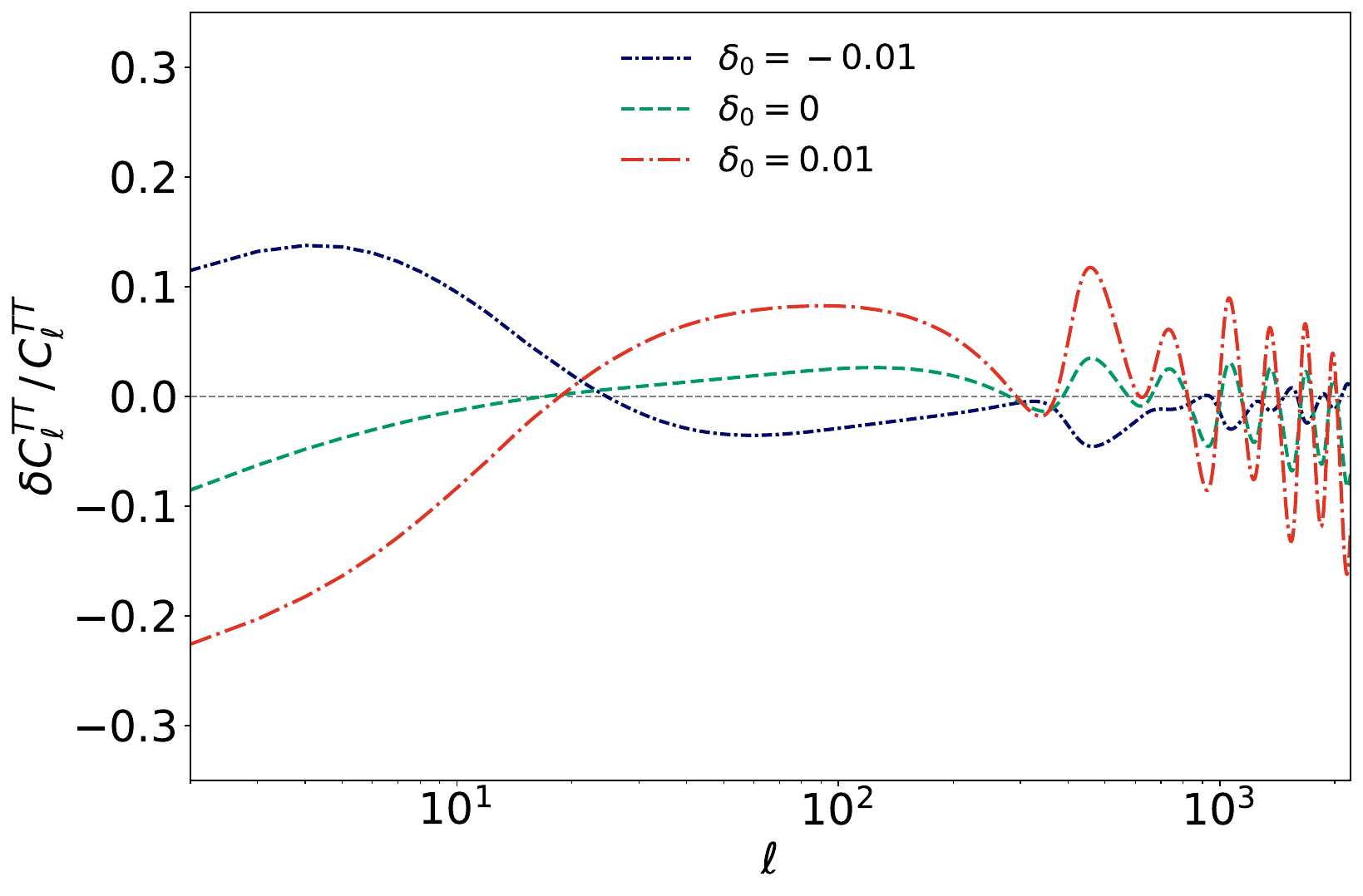}
	\includegraphics[width=0.47\textwidth]{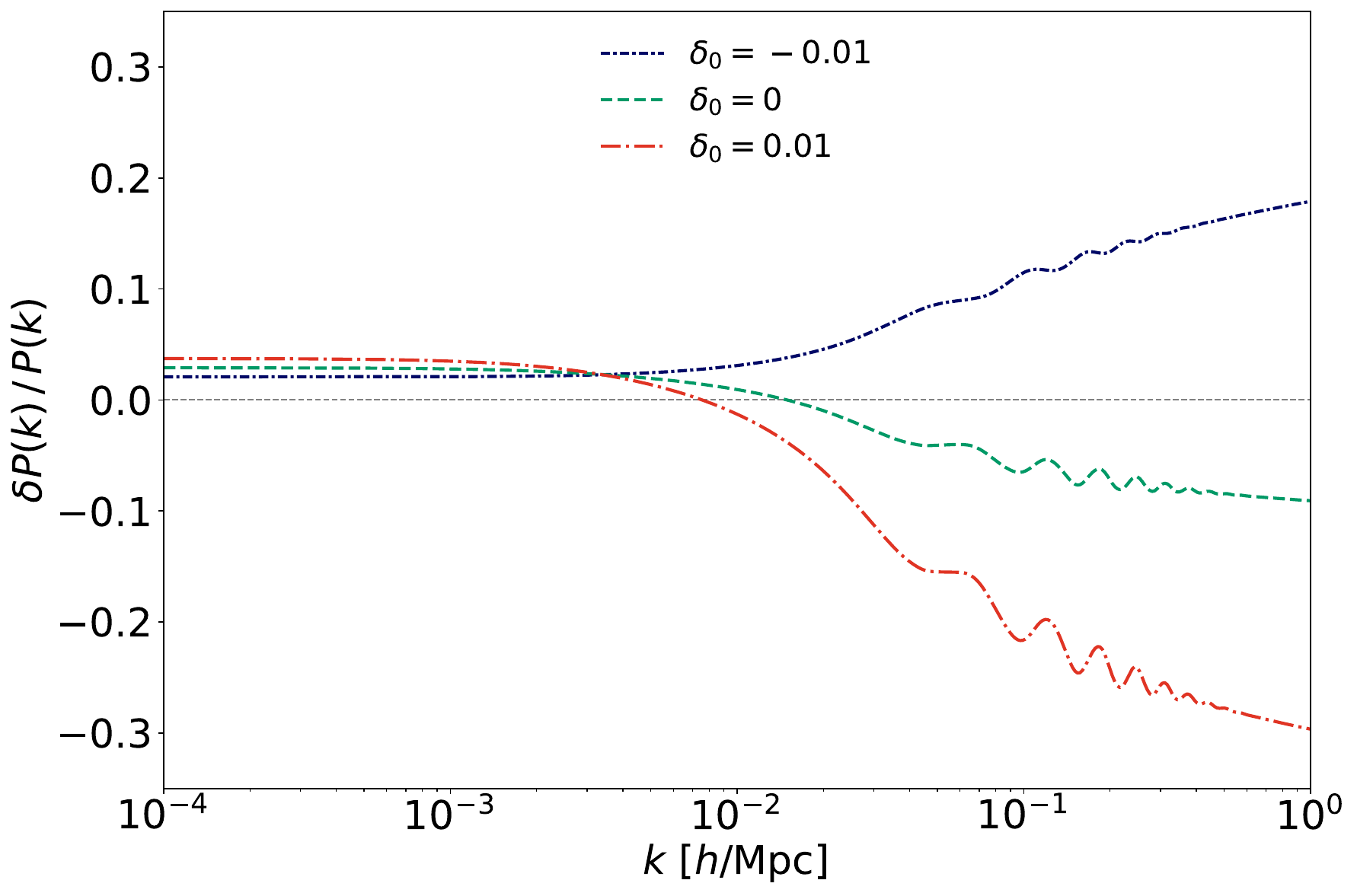}
\caption{\textit{Residuals of the CMB TT angular power spectrum (left panel) and
matter power spectrum (right panel) for the {\bf IDE$+w$} scenario, relative to
the Planck 2018 best-fit $\Lambda$CDM model~\cite{Planck:2018vyg}, for different
values of the interaction parameter $\delta_0$. The dark energy equation of
state is fixed to $w_0=-0.9$. The remaining cosmological parameters are fixed
to the mean values obtained from the CMB+DESI+PantheonPlus analysis (fourth
column of Table~\ref{table-1}).}}
	\label{fig:wde-delta0-spectra}
\end{figure*}

\begin{figure*}
	\centering
	\includegraphics[width=0.47\textwidth]{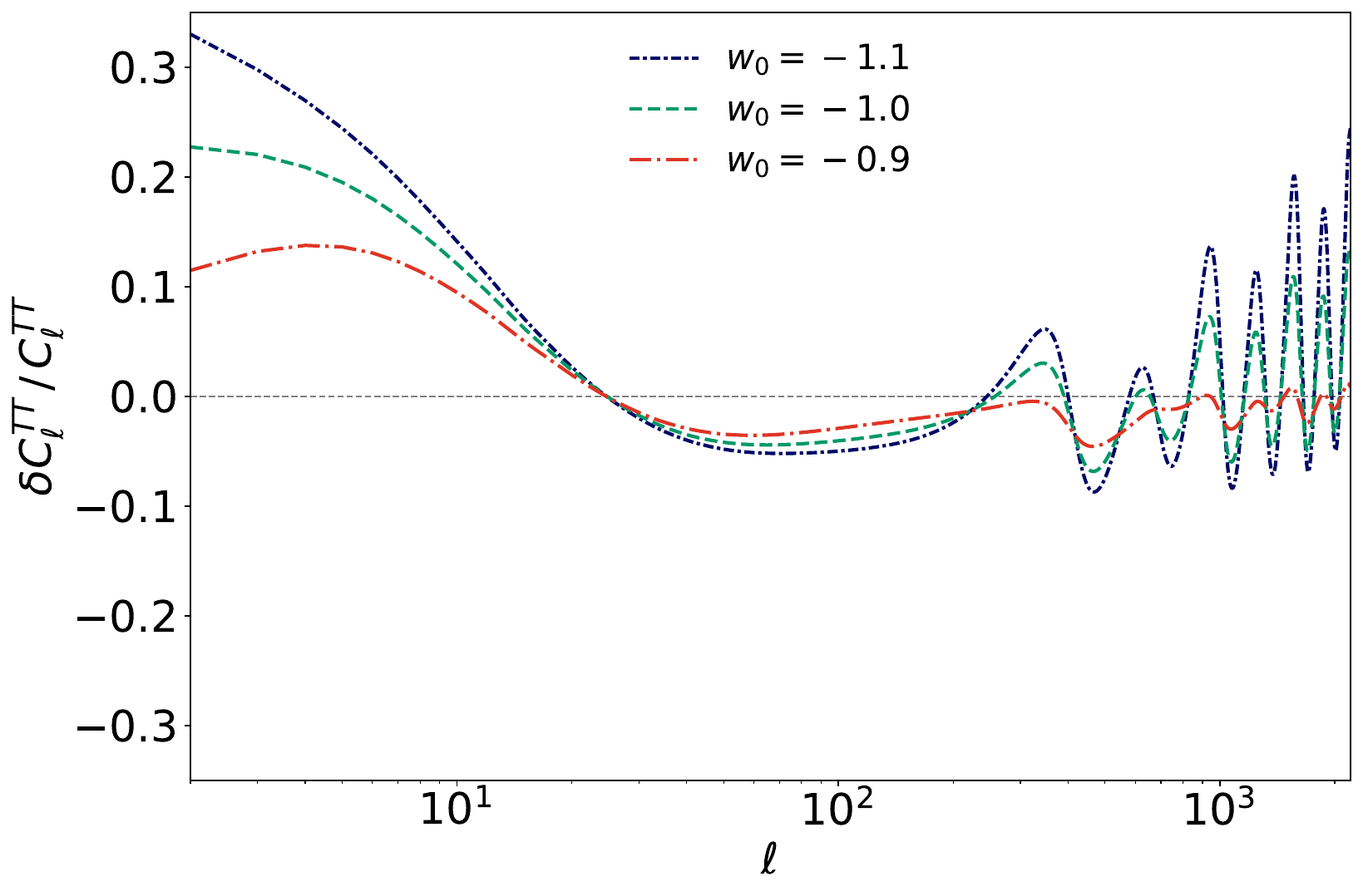}
	\includegraphics[width=0.47\textwidth]{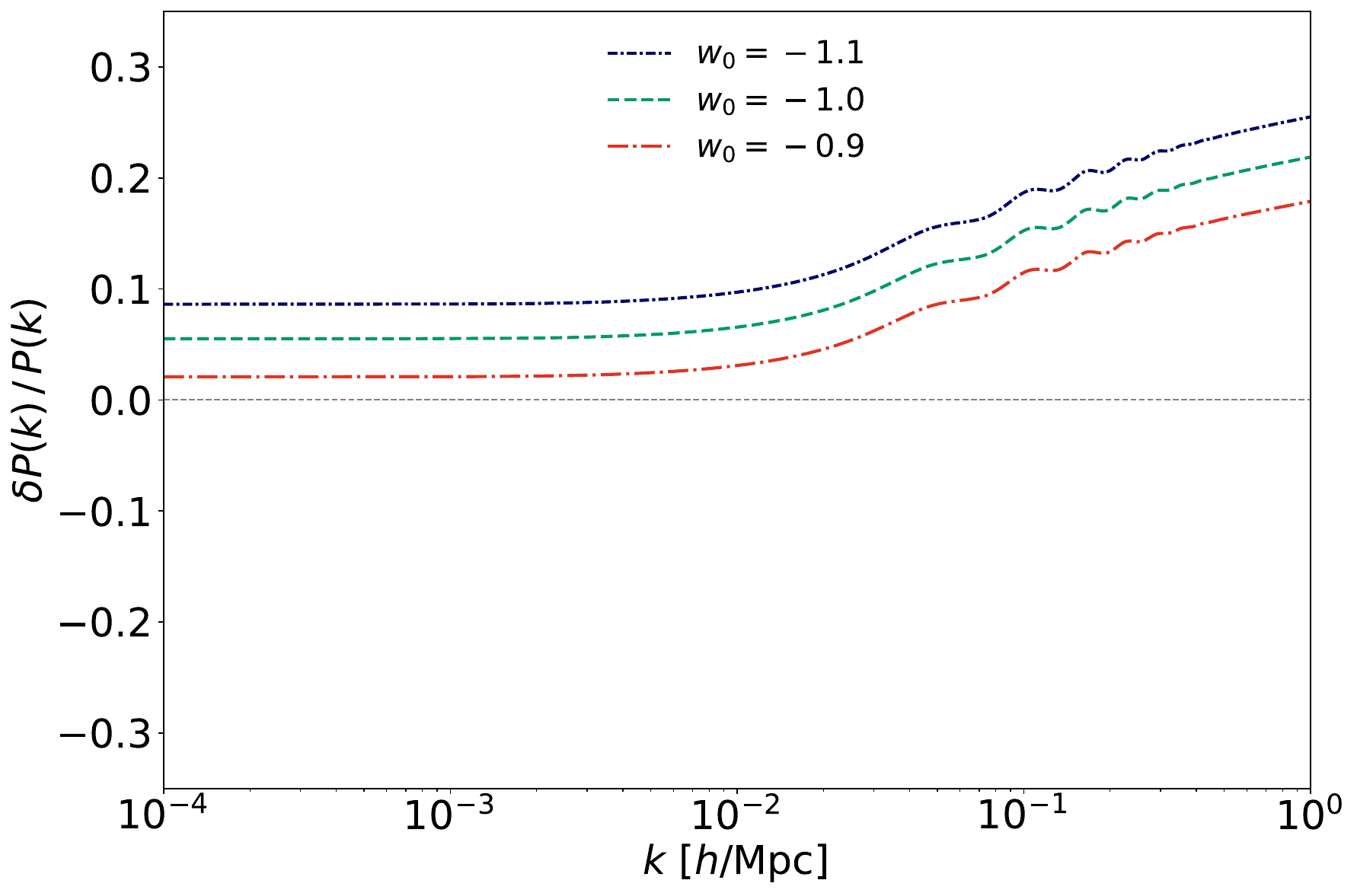}
\caption{\textit{Residuals of the CMB TT angular power spectrum (left panel) and
matter power spectrum (right panel) for the {\bf IDE$+w$} scenario, relative to
the Planck 2018 best-fit $\Lambda$CDM model~\cite{Planck:2018vyg}, for different
values of the dark energy equation-of-state parameter $w_0$. The interaction
parameter is fixed to $\delta_0=-0.01$. The remaining cosmological parameters
are fixed to the mean values obtained from the CMB+DESI+PantheonPlus analysis
(fourth column of Table~\ref{table-1}).}}
	\label{fig:wde-w0-spectra}
\end{figure*}

\section{Observational constraints}
\label{sec-results}

\begin{figure*}
	\includegraphics[width=0.85\textwidth]{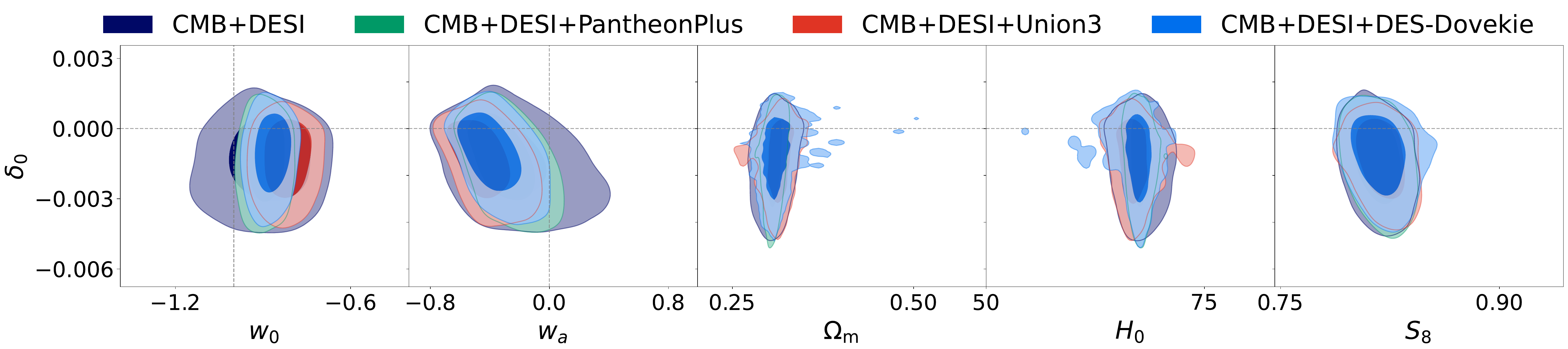}
	\includegraphics[width=0.85\textwidth]{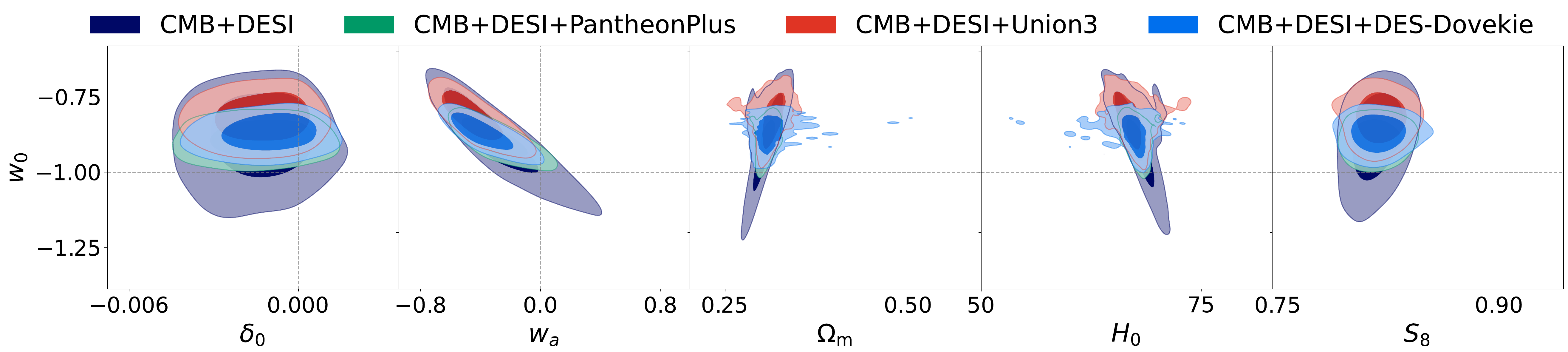}
	\includegraphics[width=0.85\textwidth]{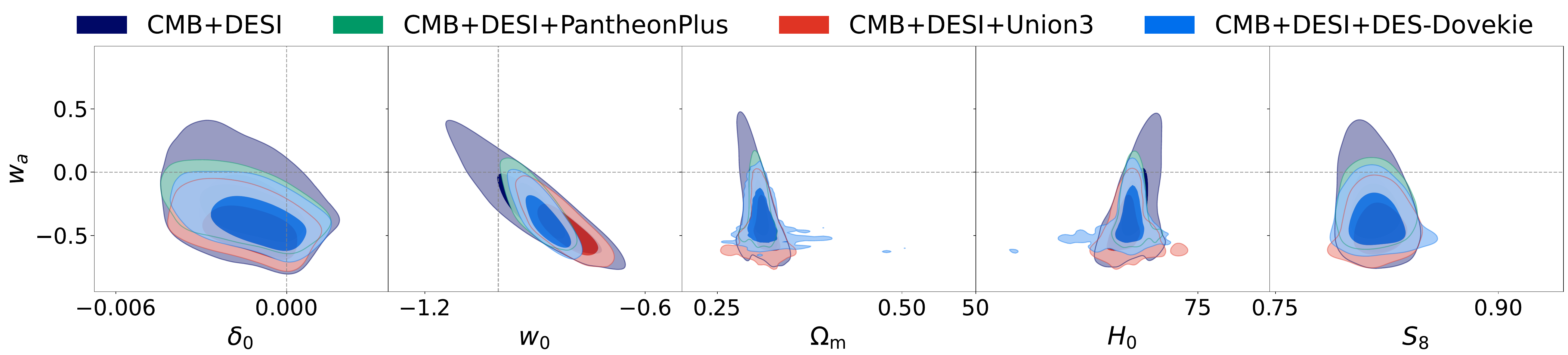}			
	\caption{\textit{Marginalized one-dimensional posterior distributions and
two-dimensional confidence contours for the main cosmological parameters of
the {\bf IDE$+w_0w_a$} scenario. Results are displayed for the CMB, CMB+DESI,
CMB+DESI+PantheonPlus, CMB+DESI+Union3, and CMB+DESI+DES-Dovekie dataset
combinations.}}
	\label{contour-2}
\end{figure*}

\begin{figure*}
	\centering
	\includegraphics[width=0.47\textwidth]{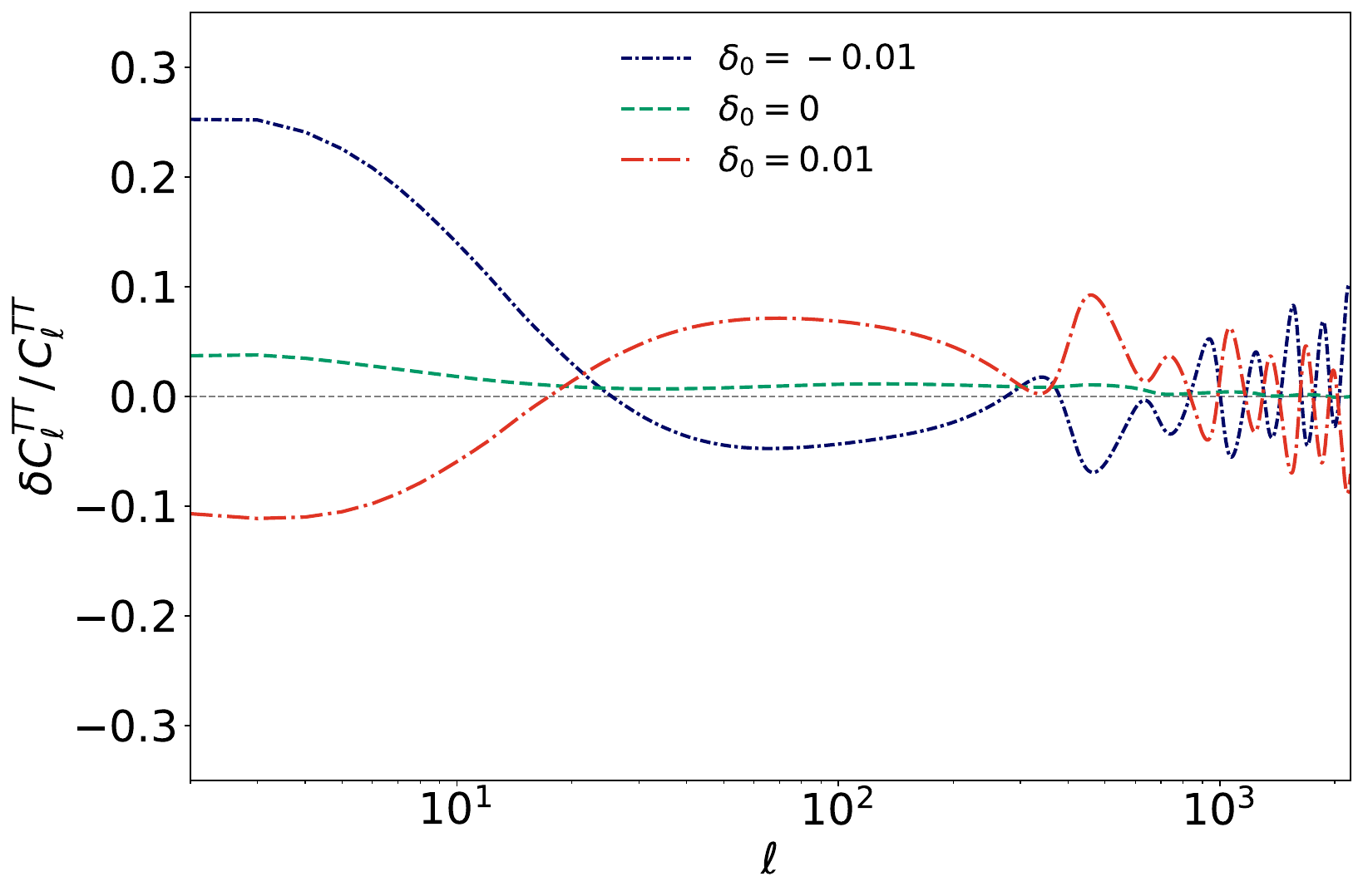}
	\includegraphics[width=0.47\textwidth]{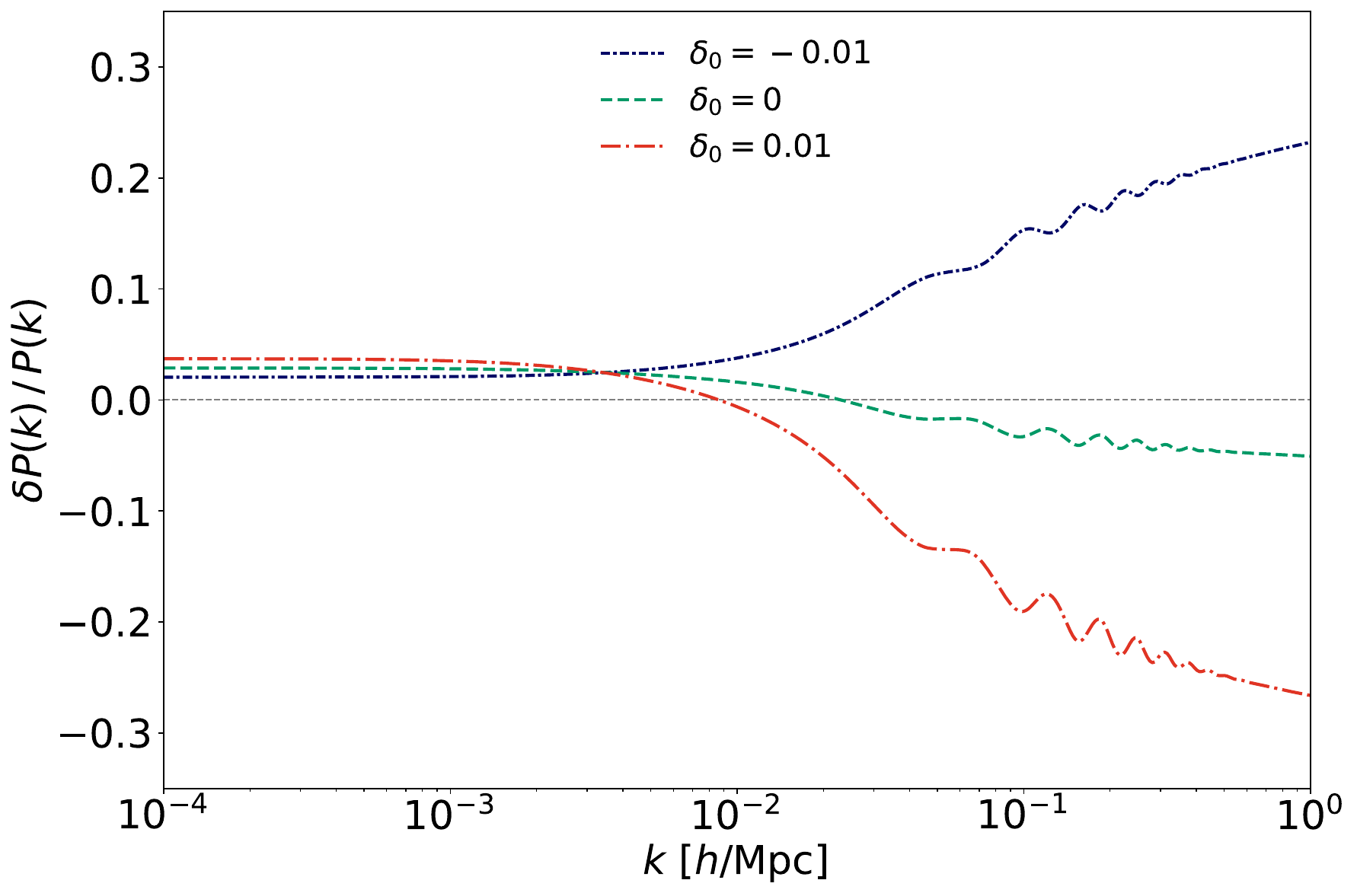}
\caption{\textit{Residuals of the CMB TT angular power spectrum (left panel) and
matter power spectrum (right panel) for the {\bf IDE$+w_0w_a$} scenario,
relative to the Planck 2018 best-fit $\Lambda$CDM model~\cite{Planck:2018vyg},
for different values of the interaction parameter $\delta_0$. The dark energy
equation-of-state parameters are fixed to $w_0=-0.9$ and $w_a=-0.3$. The
remaining cosmological parameters are fixed to the mean values obtained from the
CMB+DESI+PantheonPlus analysis (fourth column of Table~\ref{table-2}).}}
	\label{fig:cpl-delta0-spectra}
\end{figure*}

\begin{figure*}
	\centering
	\includegraphics[width=0.47\textwidth]{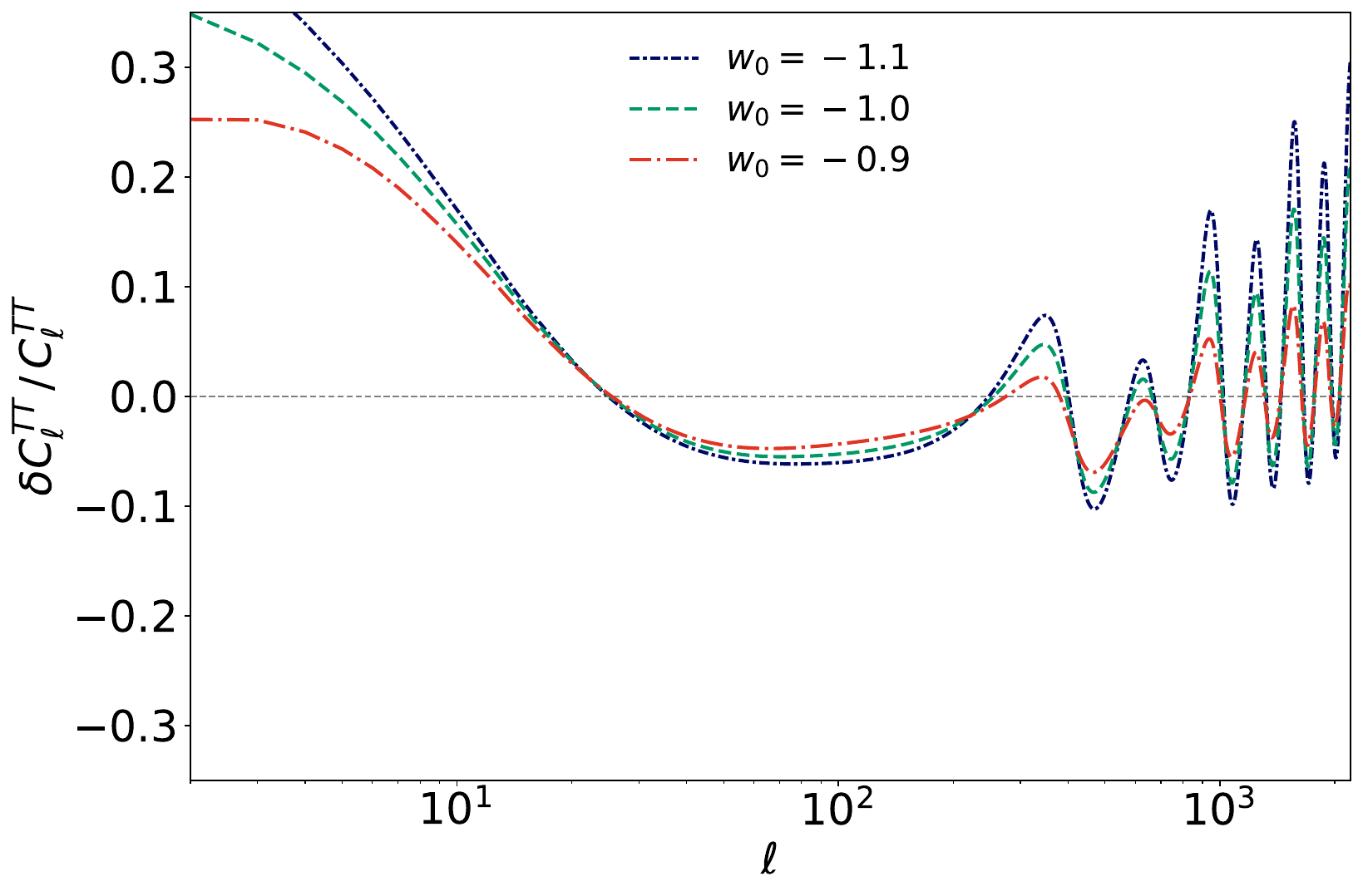}
	\includegraphics[width=0.47\textwidth]{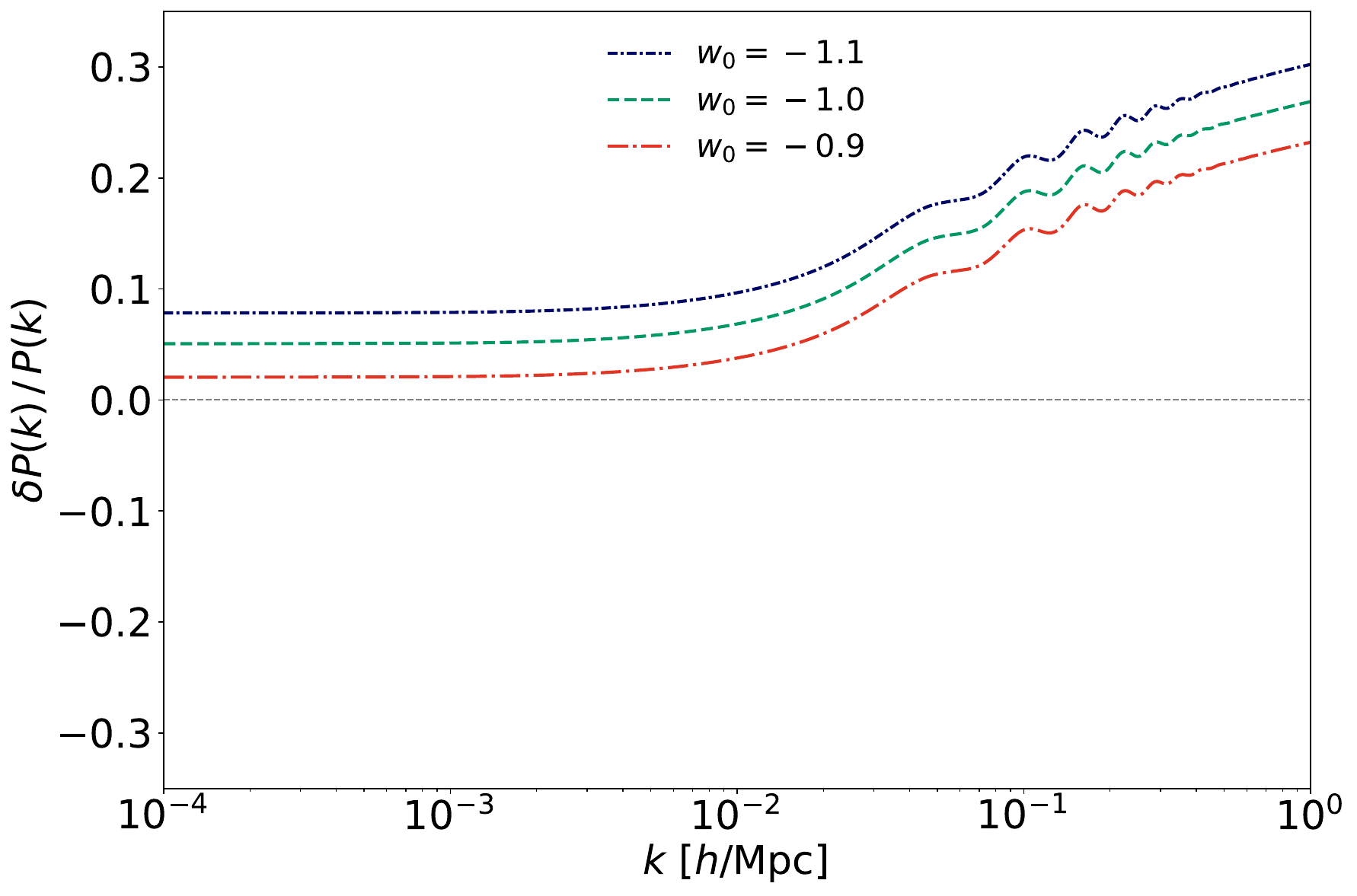}
\caption{\textit{Residuals of the CMB TT angular power spectrum (left panel) and
matter power spectrum (right panel) for the {\bf IDE$+w_0w_a$} scenario,
relative to the Planck 2018 best-fit $\Lambda$CDM model~\cite{Planck:2018vyg},
for different values of the dark energy equation-of-state parameter $w_0$. The
interaction parameter and the CPL evolution parameter are fixed to
$\delta_0=-0.01$ and $w_a=-0.3$, respectively. The remaining cosmological
parameters are fixed to the mean values obtained from the CMB+DESI+PantheonPlus
analysis (fourth column of Table~\ref{table-2}).}}
	\label{fig:cpl-w0-spectra}
\end{figure*}

\begin{figure*}
	\centering
	\includegraphics[width=0.47\textwidth]{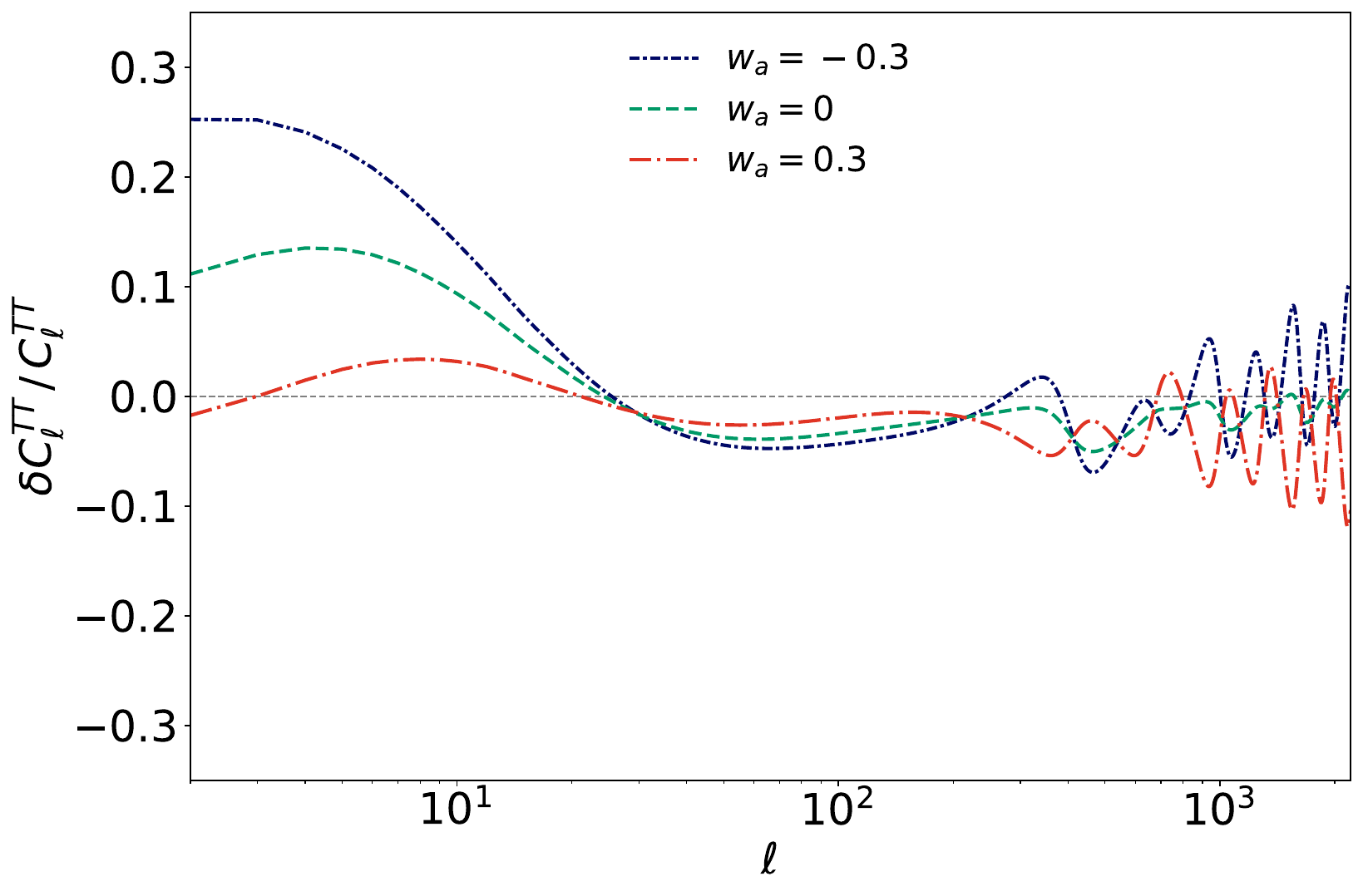}
	\includegraphics[width=0.47\textwidth]{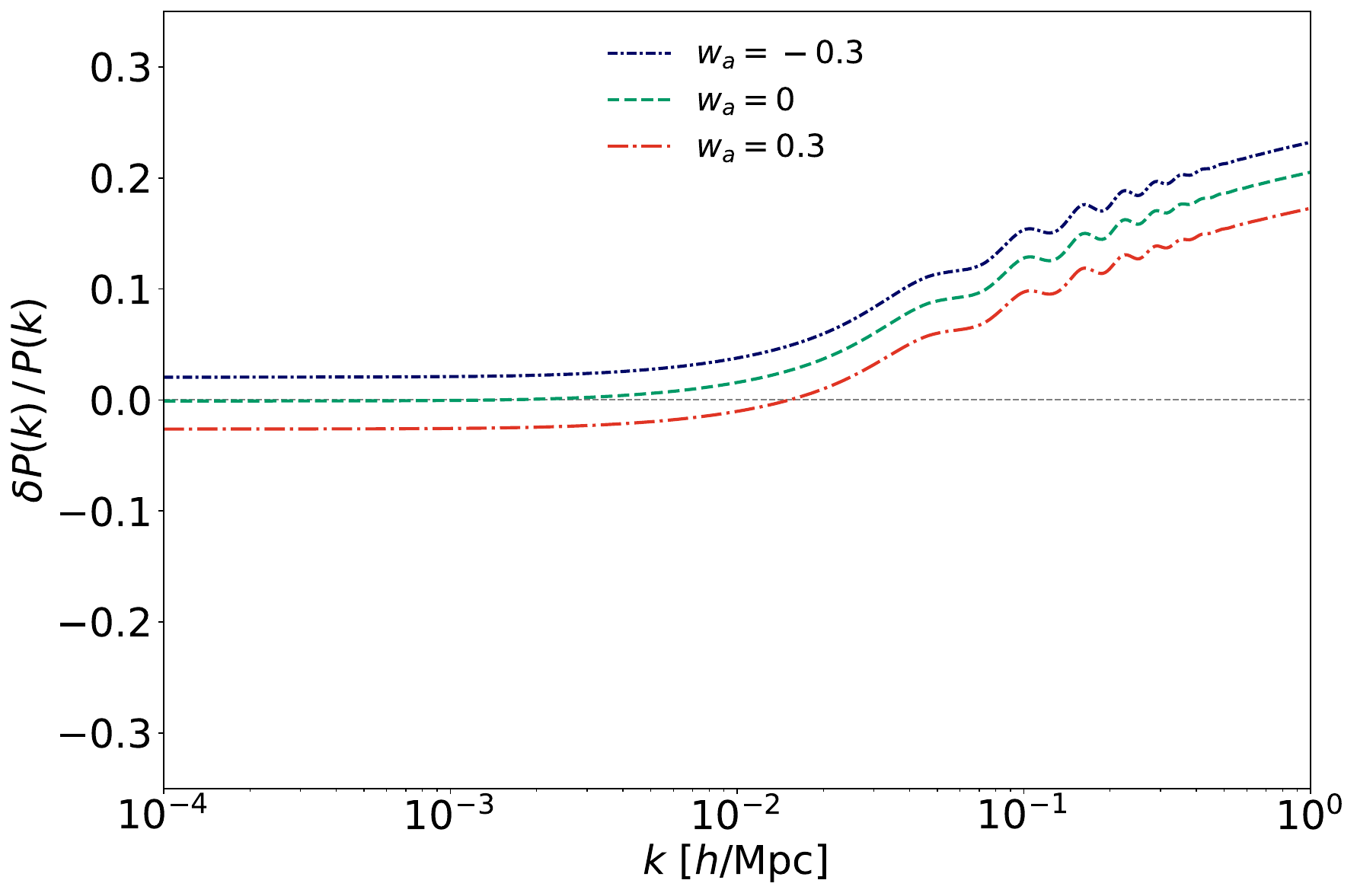}
\caption{\textit{Residuals of the CMB TT angular power spectrum (left panel) and
matter power spectrum (right panel) for the {\bf IDE$+w_0w_a$} scenario,
relative to the Planck 2018 best-fit $\Lambda$CDM model~\cite{Planck:2018vyg},
for different values of the CPL evolution parameter $w_a$. The interaction
parameter and the present-day dark energy equation-of-state parameter are fixed
to $\delta_0=-0.01$ and $w_0=-0.9$, respectively. The remaining cosmological
parameters are fixed to the mean values obtained from the CMB+DESI+PantheonPlus
analysis (fourth column of Table~\ref{table-2}).}}
	\label{fig:cpl-wa-spectra}
\end{figure*}

In this section we present the key results obtained for the present 
interacting scenarios. We consider several dataset combinations in order to 
robustly constrain the interacting scenarios. In total, we perform five 
analyses: CMB alone, CMB+DESI, and CMB+DESI+SNIa, with the latter considered 
separately for the PantheonPlus, Union3, and DES-Dovekie compilations.     
Throughout the analysis, the total neutrino mass is 
fixed to
$\sum m_\nu=0.06$ eV and the effective number of relativistic species is fixed 
to
$N_{\rm eff}=3.04$.    In Table \ref{table-1}   we present  
the 68\% and 95\% CL constraints on the free and derived parameters of the  
{\bf IDE$+w$} 
interacting scenario,  and in Figs. 
\ref{contour-1}-\ref{fig:wde-w0-spectra} we display the corresponding plots. 
Similarly, in  Table \ref{table-2} and in Figs.  
\ref{contour-2}-\ref{fig:cpl-wa-spectra} we present the results for the  
{\bf IDE$+w_0w_a$} scenario.

\begingroup
\squeezetable                                   
\begin{center}  
	\begin{table*}
	\caption{68\% and 95\% confidence level constraints on the free and derived
cosmological parameters of the {\bf IDE$+w_0w_a$} scenario for the CMB,
CMB+DESI, CMB+DESI+PantheonPlus, CMB+DESI+Union3, and
CMB+DESI+DES-Dovekie dataset combinations.}
		\begin{tabular}{cccccc}
			\hline\hline
			Parameters & CMB & CMB+DESI & CMB+DESI+PantheonPlus & 
CMB+DESI+Union3 & CMB+DESI+DES-Dovekie  \\ \hline
			
			$\Omega_\mathrm{b} h^2$ & 
$0.02242_{-0.00018-0.00033}^{+0.00016+0.00035}$ & 
$0.02237_{-0.00016-0.00031}^{+0.00015+0.00031}$ & 
$0.02236_{-0.00016-0.00033}^{+0.00016+0.00031}$ & 
$0.02238_{-0.00016-0.00030}^{+0.00016+0.00032}$ & 
$0.02238_{-0.00017-0.00031}^{+0.00016+0.00031}$ \\
			
			$\Omega_\mathrm{dm} h^2$ & 
$0.12716_{-0.00490-0.00850}^{+0.00305+0.00970}$ & 
$0.11753_{-0.00094-0.00211}^{+0.00113+0.00202}$ & 
$0.11761_{-0.00099-0.00213}^{+0.00110+0.00212}$ & 
$0.11754_{-0.00093-0.00204}^{+0.00102+0.00185}$ & 
$0.11779_{-0.00084-0.00193}^{+0.00104+0.00183}$ \\
			
			$100\theta_\mathrm{MC}$ & 
$1.04014_{-0.00042-0.00097}^{+0.00047+0.00086}$ & 
$1.04087_{-0.00045-0.00236}^{+0.00055+0.00186}$ & 
$1.04086_{-0.00035-0.00170}^{+0.00047+0.00125}$ & 
$1.04082_{-0.00064-0.00268}^{+0.00082+0.00226}$ & 
$1.04000_{--0.00009-0.00838}^{+0.00165+0.00338}$ \\
			
			$\tau$ & $0.0520_{-0.0075-0.0149}^{+0.0075+0.0157}$ & 
$0.0572_{-0.0085-0.0145}^{+0.0071+0.0167}$ & 
$0.0570_{-0.0082-0.0145}^{+0.0073+0.0157}$ & 
$0.0572_{-0.0080-0.0144}^{+0.0070+0.0157}$ & 
$0.0573_{-0.0077-0.0153}^{+0.0074+0.0158}$ \\
			
			$n_\mathrm{s}$ & $0.9660_{-0.0053-0.0109}^{+0.0054+0.0105}$ & 
$0.9773_{-0.0034-0.0069}^{+0.0035+0.0070}$ & 
$0.9771_{-0.0035-0.0068}^{+0.0035+0.0070}$ & 
$0.9773_{-0.0036-0.0069}^{+0.0035+0.0067}$ & 
$0.9771_{-0.0034-0.0068}^{+0.0034+0.0069}$ \\
			
			$\ln(10^{10} A_\mathrm{s})$ & $3.051_{-0.016-0.031}^{+0.016+0.032}$ 
& $3.057_{-0.017-0.031}^{+0.015+0.034}$ & $3.057_{-0.017-0.031}^{+0.015+0.032}$ 
& $3.057_{-0.017-0.030}^{+0.015+0.033}$ & $3.058_{-0.016-0.032}^{+0.016+0.032}$ 
\\
			
			$\delta_0$ & $0.0060_{-0.0045-0.0074}^{+0.0023+0.0094}$ & 
$-0.0013_{-0.0010-0.0026}^{+0.0014+0.0022}$ & 
$-0.0013_{-0.0010-0.0025}^{+0.0015+0.0023}$ & 
$-0.0013_{-0.0009-0.0023}^{+0.0014+0.0020}$ & 
$-0.0011_{-0.0009-0.0024}^{+0.0013+0.0021}$ \\
			
			$w_0$ & $-1.251_{-0.551-1.152}^{+0.519+1.152}$ & 
$-0.883_{-0.080-0.203}^{+0.110+0.186}$ & $-0.891_{-0.041-0.087}^{+0.045+0.081}$ 
& $-0.819_{-0.052-0.113}^{+0.057+0.103}$ & 
$-0.870_{-0.037-0.086}^{+0.044+0.076}$ \\
			
			$w_a$ & $0.545_{-0.752-1.544}^{+0.820+1.529}$ & 
$-0.310_{-0.273-0.402}^{+0.140+0.503}$ & $-0.302_{-0.177-0.267}^{+0.110+0.314}$ 
& $-0.444_{-0.165-0.260}^{+0.102+0.307}$ & 
$-0.380_{-0.172-0.251}^{+0.100+0.306}$ \\
			\hline 
            
			$\Omega_\mathrm{m}$ & $0.3673_{-0.1266-0.1848}^{+0.0636+0.2292}$ & 
$0.3078_{-0.0133-0.0273}^{+0.0132+0.0297}$ & 
$0.3074_{-0.0066-0.0184}^{+0.0068+0.0156}$ & 
$0.3140_{-0.0096-0.0412}^{+0.0108+0.0276}$ & 
$0.3165_{-0.0166-0.0292}^{+0.0013+0.0560}$ \\
			
			$\sigma_8$ & $0.801_{-0.065-0.127}^{+0.063+0.125}$ & 
$0.807_{-0.016-0.033}^{+0.016+0.033}$ & $0.807_{-0.013-0.024}^{+0.011+0.027}$ & 
$0.801_{-0.015-0.033}^{+0.014+0.040}$ & $0.801_{-0.012-0.047}^{+0.017+0.034}$ \\
			
			$H_0$ [km/s/Mpc] & $65.95_{-9.71-17.89}^{+8.39+18.10}$ & 
$67.64_{-1.50-2.98}^{+1.46+3.10}$ & $67.66_{-0.77-1.67}^{+0.63+1.63}$ & 
$66.98_{-1.29-3.28}^{+0.90+3.62}$ & $67.10_{-0.54-5.58}^{+1.27+3.06}$ \\
			
			$S_8$ & $0.867_{-0.063-0.105}^{+0.047+0.128}$ & 
$0.817_{-0.011-0.022}^{+0.011+0.023}$ & $0.817_{-0.011-0.021}^{+0.011+0.023}$ & 
$0.819_{-0.011-0.024}^{+0.012+0.022}$ & $0.819_{-0.013-0.027}^{+0.011+0.026}$ \\
			
			$r_{\rm{drag}}$ [Mpc] & $146.61_{-0.39-0.82}^{+0.39+0.77}$ & 
$147.50_{-0.23-0.42}^{+0.21+0.46}$ & $147.49_{-0.23-0.45}^{+0.23+0.45}$ & 
$147.49_{-0.23-0.44}^{+0.22+0.45}$ & $147.48_{-0.20-0.41}^{+0.20+0.43}$ \\
			
			\hline
			
			$w_0~(\sigma)$ & $0.4991$ & $1.1701$ & $2.4336$ & $3.0468$ & 
$2.8558$\\
			
			$w_a~(\sigma)$ & $0.6889$ & $1.2654$ & $1.7488$ & $2.5465$ & 
$2.1752$\\
			
			$\delta_0~(\sigma)$ & $1.8682$ & $1.0885$ & $1.0657$ & $1.2417$ & 
$0.9482$\\

			$\rm{ln}\mathcal{B}_{ij}$ & $-8.4$ & $-11.5$ & $-11.5$ & $-8.0$ & 
$-9.3$\\
			
			\hline                                                      
		\end{tabular}                                                       
			\label{table-2}                   
	\end{table*}                                     
\end{center}
\endgroup

\subsection{Interacting dark energy with constant equation of state}

We first consider the case in which the dark energy equation of state is 
constant but
not fixed to the cosmological constant value, namely the {\bf IDE+$w$} 
scenario. The
constraints are summarized in Table~\ref{table-1}, while the corresponding 
posterior
distributions and parameter correlations are shown in Fig.~\ref{contour-1}. The
interaction parameter $\delta_0$ and the present-day dark energy equation of 
state
$w_0$ represent the two key parameters controlling the departure from the 
standard
$\Lambda$CDM scenario.

We begin with the CMB-only constraints. In this case, a mild preference for a
non-vanishing interaction is observed, with
\[
\delta_0=0.0043^{+0.0024}_{-0.0026},
\]
corresponding to a deviation from zero slightly above the $1\sigma$ level. Since
$\delta_0>0$, the preferred direction of the energy transfer is from dark 
energy 
to
dark matter. This leads to an increased matter abundance,
$\Omega_m=0.396^{+0.041}_{-0.101}$, which in turn shifts the inferred Hubble 
constant
towards lower values,
\[
H_0=62.18^{+7.15}_{-4.74}\ {\rm km/s/Mpc},
\]
while increasing the clustering amplitude,
\[
S_8=0.882^{+0.034}_{-0.059}.
\]
The dark energy equation of state is found to lie in the quintessence region, 
while the
Bayesian evidence still strongly favors $\Lambda$CDM over this extended 
scenario.
Therefore, although the CMB data alone allow a mild indication of interaction, 
they do
not provide statistical support for introducing an additional dark-sector 
degree 
of
freedom.

The inclusion of DESI BAO measurements significantly changes the preferred 
region of
the interaction parameter. The posterior shifts from positive to negative 
values,
\[
\delta_0=-0.0018^{+0.0013}_{-0.0013},
\]
indicating a mild preference for energy transfer from dark matter to dark 
energy.
At the same time, the dark energy equation of state becomes fully consistent 
with the
cosmological constant boundary,
\[
w_0=-0.997\pm0.046.
\]
The reduced matter density leads to an increase of the inferred expansion rate,
with
\[
H_0=68.69^{+0.97}_{-1.01}\ {\rm km/s/Mpc},
\]
bringing the model closer to the standard low-redshift expansion history. 
However,
despite the improved phenomenological behavior, the Bayesian evidence remains
unfavorable compared with $\Lambda$CDM, with
$\ln\mathcal{B}_{ij}=-8.6$.

The addition of Type Ia supernovae strengthens the preference for a negative
interaction. For both CMB+DESI+PantheonPlus and CMB+DESI+Union3 combinations, 
the
interaction parameter deviates from zero at more than $2\sigma$,
\[
\delta_0=-0.0025^{+0.0012}_{-0.0012},
\]
and
\[
\delta_0=-0.0028^{+0.0012}_{-0.0013},
\]
respectively. Thus, the combined low-redshift observations consistently favor a 
mild
energy transfer from dark matter to dark energy. Nevertheless, the Bayesian 
evidence
continues to penalize the additional interaction degree of freedom, indicating 
that
the improvement in the goodness of fit is not sufficient to establish a 
departure from
$\Lambda$CDM.

In summary, the constant-EoS interacting scenario reveals an interesting 
interplay
between the nature of dark energy and the dark-sector coupling: once 
low-redshift
distance measurements are included, the data favor a small negative interaction,
while the evidence remains statistically inconclusive. This demonstrates that a
constant non-$\Lambda$ dark energy component can alter the preferred interaction
behavior, but does not yet provide decisive evidence for physics beyond the 
standard
cosmological model.


We finally investigate the characteristic signatures of the {\bf IDE+$w$} 
scenario in 
the
CMB temperature anisotropies and matter power spectrum. 
Figs. \ref{fig:wde-delta0-spectra}
and \ref{fig:wde-w0-spectra} show the residuals with respect to the reference
$\Lambda$CDM cosmology for different values of $\delta_0$ and $w_0$.

Fig. \ref{fig:wde-delta0-spectra} illustrates the impact of the dark-sector
interaction. A non-zero $\delta_0$ modifies the dark matter dilution history 
and,
consequently, the growth of cosmic structures. Positive values of $\delta_0$
correspond to a slower dilution of dark matter due to energy transfer from dark
energy to dark matter, enhancing clustering and increasing the deviations in the
matter power spectrum. Conversely, negative values lead to faster dark matter
dilution and suppress structure formation. The modified background evolution and
gravitational potentials also leave imprints on the CMB TT spectrum.

Fig. \ref{fig:wde-w0-spectra} shows the effect of varying $w_0$ while keeping
$\delta_0$ fixed. Changes in the dark energy equation of state modify the 
expansion
history and the evolution of gravitational potentials, affecting both CMB
anisotropies through late-time effects such as the integrated Sachs--Wolfe 
effect
and lensing, and the growth of matter perturbations. On large scales, the matter
power spectrum residuals approach an approximately constant value, reflecting 
the
dominant impact on the overall growth amplitude in the linear regime, while 
smaller
scales exhibit a stronger dependence on the detailed perturbation evolution.

These results demonstrate that the interaction parameter and the dark energy 
equation
of state leave distinct signatures in cosmological observables, connecting the
parameter constraints with the underlying dark-sector physics.


\subsection{Interacting dark energy with dynamical equation of state}

The constraints for the {\bf IDE+$w_0w_a$} scenario are summarized in 
Table~\ref{table-2},
while the corresponding posterior distributions and parameter correlations are 
shown
in Fig.~\ref{contour-2}. The three panels of Fig.~\ref{contour-2} display the
correlations among the main parameters of this scenario, namely the interaction
parameter $\delta_0$ and the dark energy equation-of-state parameters $w_0$ and 
$w_a$.

We first consider the CMB-only constraints, shown in the second column of
Table~\ref{table-2}. A mild preference for a non-zero interaction is found, with
$\delta_0\neq0$ at approximately $1.87\sigma$. The preferred positive value of
$\delta_0$ corresponds to an energy transfer from dark energy to dark matter, 
leading
to an increased matter density,
\[
\Omega_m=0.3673^{+0.0636}_{-0.1266},
\]
and consequently to a lower value of $H_0$ due to the well-known 
anti-correlation
between these two parameters. The present-day dark energy equation of state is
shifted towards the phantom region,
\[
w_0=-1.251^{+0.519}_{-0.551},
\]
although the cosmological constant value remains within the $1\sigma$ region.
Similarly, $w_a$ is consistent with zero. Therefore, CMB data alone allow a mild
interaction preference, but do not provide evidence for a departure from a
cosmological constant dark energy sector.

The inclusion of DESI BAO data modifies this picture. The indication for a 
non-zero
interaction becomes weaker, with $\delta_0\neq0$ only at approximately the
$1\sigma$ level. At the same time, the dark energy equation of state shifts 
towards
the quintessence region, with a mild indication of dynamical behavior. The 
remaining
cosmological parameters, including $H_0$, $\Omega_m$, and $S_8$, remain 
compatible
with their $\Lambda$CDM values.

When the three supernova compilations are further combined with CMB+DESI, the
constraints become consistent across the different datasets. The interaction
parameter remains only mildly displaced from zero, with a significance ranging 
from
approximately $0.95\sigma$ to $1.24\sigma$. In contrast, the preference for a
quintessence-like present-day equation of state becomes stronger, with $w_0>-1$
favored at the level of approximately $2.43\sigma-3.05\sigma$. Moreover, the
evolution parameter $w_a$ shows a preference for non-zero values at 
approximately
$1.75\sigma-2.55\sigma$, indicating support for a dynamical dark energy 
component.
The remaining cosmological parameters continue to show values close to those of
$\Lambda$CDM.

Summarizing, when the dark energy equation of state is allowed to evolve, the 
DESI-era
datasets primarily favor dynamical dark energy rather than an additional 
dark-sector
interaction. The interaction parameter remains compatible with zero, suggesting 
that
the evidence for departures from $\Lambda$CDM is mainly associated with the 
evolution
of the dark energy equation of state.

We finally investigate the characteristic signatures of the {\bf IDE+$w_0w_a$} 
scenario in
the CMB temperature anisotropies and matter power spectrum. Figures~
\ref{fig:cpl-delta0-spectra}, \ref{fig:cpl-w0-spectra}, and
\ref{fig:cpl-wa-spectra} show the residuals with respect to $\Lambda$CDM when
varying $\delta_0$, $w_0$, and $w_a$, respectively. The interaction parameter
directly affects the dark matter dilution history and therefore the growth of
structures, while $w_0$ and $w_a$ modify the expansion history and the 
evolution 
of
gravitational potentials. Consequently, all three parameters leave distinct
signatures in the CMB TT spectrum and matter power spectrum. As in the 
constant-EoS
case, the matter power spectrum residuals approach an approximately constant 
value
on large scales, reflecting the dominant effect of these parameters on the 
overall
growth amplitude in the linear regime.

\section{Summary and conclusions}
\label{sec-summary}

The possibility that dark matter and dark energy interact through a 
non-gravitational
exchange represents one of the most interesting extensions of the standard
$\Lambda$CDM cosmological model. In such scenarios, the standard dilution law of
pressureless dark matter, $\rho_{\rm dm}\propto a^{-3}$, is modified. In this 
work,
we have considered the phenomenological parametrization
\[
\rho_{\rm dm}\propto a^{-3+\delta_0},
\]
where the constant parameter $\delta_0$ quantifies the strength of the 
dark-sector
interaction, with $\delta_0=0$ recovering the standard non-interacting case.

Motivated by the recent DESI DR2 indications for a possible departure from a 
pure
cosmological constant, we have investigated how the nature of dark energy 
affects the
inferred strength of dark-sector interactions. In particular, we have extended 
the
usual interacting dark energy framework by considering two different dark energy
descriptions: a constant equation of state, $w(a)=w_0$, where $w_0$ is not 
fixed 
to
$-1$, and a dynamical equation of state described by the CPL parametrization,
$w(a)=w_0+w_a(1-a)$. These two scenarios allow us to explore whether the 
possible
signatures of dark-sector interactions depend on the intrinsic properties of 
dark
energy.

We have constrained these models using a combination of current cosmological
observations, including Planck 2018 CMB temperature and polarization spectra, 
DESI DR2
BAO measurements, and three independent Type Ia supernova compilations:
PantheonPlus, Union3, and DES-Dovekie. The analysis has been performed at both 
the
background and perturbation levels, allowing us to consistently investigate the
impact of the interaction on the expansion history, CMB anisotropies, and 
structure
formation.

For the constant equation-of-state interacting scenario ({\bf IDE+$w$}), the 
combined
CMB+DESI+SNIa datasets consistently favor a small negative interaction 
parameter,
corresponding to an energy transfer from dark matter to dark energy, with a
significance of approximately $2.2\sigma$. At the same time, the present-day 
dark
energy equation of state shows a preference for quintessence-like values. These
results indicate that once the dark energy sector is allowed to move away from a
strict cosmological constant, the observational preference for interaction can 
be
modified and mildly enhanced. Nevertheless, Bayesian evidence calculations show 
that
the additional degrees of freedom introduced by the interacting scenario are not
statistically favored over $\Lambda$CDM.

When the dark energy equation of state is allowed to evolve dynamically
({\bf IDE+$w_0w_a$}), the picture changes. The evidence for a non-zero 
interaction 
becomes
weak across all datasets, while the data show a moderate preference for a
quintessence-like and evolving dark energy component. For the combined
CMB+DESI+SNIa datasets, the preference for $w_0>-1$ reaches approximately
$2.4\sigma-3.1\sigma$, while the indication for a non-zero $w_a$ reaches
approximately $1.7\sigma-2.5\sigma$. Therefore, within this more general 
framework,
the DESI-era preference appears to be primarily associated with the dynamical 
nature
of dark energy rather than with an additional interaction between the dark 
fluids.

We have also studied the characteristic signatures of the interaction and dark 
energy
parameters in the CMB TT angular power spectrum and matter power spectrum. The
interaction parameter modifies the dark matter evolution and consequently the 
growth
of structures, while the dark energy equation-of-state parameters affect the
expansion history and the evolution of gravitational potentials. These effects 
leave
distinct imprints on cosmological observables, providing a direct connection 
between
the parameter constraints and the underlying dark-sector physics.

Overall, our results indicate that the interpretation of possible deviations 
from
$\Lambda$CDM depends crucially on the assumed properties of the dark energy 
sector.
A constant non-standard equation of state can strengthen the preference for
dark-sector interactions, whereas allowing a fully dynamical dark energy 
component
shifts the preference towards evolving dark energy itself. In all cases, 
however,
Bayesian model comparison continues to favor $\Lambda$CDM, showing that present
observations do not yet require additional dark-sector physics. Future 
high-precision
measurements, together with more general parametrizations of the interaction
function, such as a time-dependent $\delta(a)$, will be essential to determine
whether the hints found in current data represent new physics or residual
statistical fluctuations.

\begin{acknowledgments}
 WY has been supported by the National Natural Science Foundation of China 
under Grant Nos. 12547110 and 12175096. 
AP was partially supported by FONDECYT Grant 1240514. 
ENS acknowledges the contribution of the LISA Cosmology Working Group (CosWG). 
The authors acknowledge as well support from the COST Actions CA21136 -  
Addressing 
observational tensions in cosmology with systematics and fundamental physics 
(CosmoVerse) - CA23130, Bridging high and low energies in search of quantum 
gravity (BridgeQG) and CA21106 - COSMIC WISPers in the Dark Universe: Theory, 
astrophysics and  experiments (CosmicWISPers).
\end{acknowledgments}

\bibliography{biblio}

\end{document}